\documentstyle[12pt,epsf,colordvi]{article}

\newcommand{\kappavec}{\mbox{\boldmath $\kappa$}}
\newcommand{\pvec}{\mbox{\boldmath $p$}}
\newcommand{\qvec}{\mbox{\boldmath $q$}}
\newcommand{\be}{\begin{equation}}
\newcommand{\ee}{\end{equation}}
\newcommand{\bea}{\begin{eqnarray}}
\newcommand{\eea}{\end{eqnarray}}
\newcommand{\difrac}[2]{\frac{\displaystyle #1}{\displaystyle #2}}

\begin{document}

\begin{center}
{\Large 
{\bf Extended Superscaling of Electron Scattering from Nuclei} 
}
\end{center}

\vspace{0.1in}

\begin{center}
C. Maieron\footnote{Present address: Fachbereich Physik,
Universit\"at Rostock, D-18051 Rostock, Germany}, T. W. Donnelly \\
Center for Theoretical Physics,  Laboratory for Nuclear Science \\
and Department of Physics \\
Massachusetts Institute of Technology \\
Cambridge, Massachusetts 02139-4307, USA  \\[5mm] 
and  \\[5mm]
Ingo Sick \\
 Departement  f\"ur Physik und Astronomie,
Universit\"at Basel  \\ CH4056 Basel, Switzerland 
\end{center}

\vspace*{5mm}
\begin{center}
\begin{minipage}{12.cm}
\small
An extended study of scaling of the first and second kinds 
for inclusive electron scattering from nuclei is presented. 
Emphasis is placed on the transverse response in the 
kinematic region lying above the quasielastic peak. In 
particular, for the region in which electroproduction of
resonances is expected to be important, approximate scaling
of the second kind is observed and the modest breaking of it
is shown probably to be due to 
the role played by an inelastic version of the usual scaling variable.
\end{minipage}

\vspace*{5mm}
\today
\end{center}
\normalsize

\section{Introduction} 
\label{intro}
In recent studies \cite{don99,don99a} the concepts of scaling of the first and
second kinds and superscaling have been explored, focusing on the
region of energy loss at or below the quasielastic peak in inclusive electron
scattering from nuclei. Scaling of the first kind corresponds to the
following behavior: if the inclusive cross section is divided by the
relevant single-nucleon electromagnetic cross section ({\it i.e.},
weighted by the proton and neutron numbers, $Z$ and $N$,
respectively, and with appropriate relativistic effects included ---
see \cite{don99a} for detailed discussions), then at sufficiently high
values of the momentum transfer $q$ the result so-obtained becomes a
function of a single scaling variable and not independently of $q$ and
the energy transfer $\omega$. Various definitions for the scaling
variable exist (see, for instance, \cite{don99a}); however, when $q$
is high enough they are almost always simply functionally related in
ways that yield scaling behavior in all cases. Such $q$-independence
of the reduced response $F(q,\omega)\to F(\psi)$, where
$\psi=\psi(q,\omega)$ is the scaling variable, 
is called scaling of the first kind. Moreover,
motivated by earlier work \cite{alb88}, it has been found that when
the typical momentum scale of a given nucleus ${\bar k}$ is appropriately
incorporated in the definition of the scaling variable and the reduced
response is also appropriately scaled, $F(q,\omega)\to
f(q,\omega)\equiv {\bar k}\times F(q,\omega)$, then a second type of
scaling behavior is seen --- the result becomes independent of nuclear
species. This is called scaling of the second kind. When scaling of
both the first and second kinds occurs, one calls the phenomenon
superscaling. 

The studies undertaken recently \cite{don99,don99a} demonstrated the
quality of the scaling behavior, finding that scaling of the first
kind is reasonably good for $\psi<0$ (below the quasielastic peak),
that scaling of the second kind is excellent in this region, but that
both are violated for $\psi>0$ (above the quasielastic peak). Indeed,
the scaling of the first kind has been known for some time to be badly
violated in this latter region. The recent analyses showed that so also
is scaling of the second kind broken in the $\psi>0$ region, but much
less so. Finally, in \cite{don99a} an initial attempt was made to use
the limited information on the separate longitudinal and transverse
responses, and hence on the scaling behavior of the 
respective reduced responses. The former appears to superscale,
whereas the dominant scale-breaking effects appear to reside more in
the latter.

In the present work we pick up these ideas and extend them. We begin
by updating our analysis of the relevant inclusive electron scattering
scaling behavior in the $\psi<0$ region, 
thereby obtaining refined values for the typical
nuclear momentum scale (henceforth, as in our previous work which was
motivated by the Relativistic Fermi Gas (RFG), called the Fermi
momentum $k_F$) and for a small energy shift $E_{shift}$ included to
have the quasielastic peak occur at the place where the scaling
variable is zero. Moreover, we assess the sensitivity of the results
to variations in both $k_F$ and $E_{shift}$ to provide some idea of
how much change in one or both can be tolerated when studying the
region where $\psi>0$.

We wish to focus on the scale-breaking effects, especially in the 
$\psi>0$ region, and
accordingly we have isolated the transverse response using our
previous approach. Within uncertainties that are unfortunately not as
small as is desirable, and only for a limited range of kinematics, 
it appears that the longitudinal reduced response $f_L$ in fact superscales
and reasonably satisfies the Coulomb sum rule. We shall assume that
this is a universal behavior (for all kinematics and for all nuclei
--- that is the impact of having superscaling), shall then remove the
longitudinal contributions and thereby obtain the transverse reduced
response, $f_T$. Naturally the uncertainties in knowledge of $f_L$
propagate into corresponding uncertainites in $f_T$, and given better
L/T separations the procedure could be refined. However, it is
important to the best of our current ability to isolate the transverse
part of the inclusive response as it is the one expected to contain
the leading scale-breaking effects \cite{don99a}.

Once $f_T$ has been isolated we focus on the region above the
quasielastic peak, exploring the scaling behavior of $f_T$ as a
function of $q$ and $k_F$, namely, for first- and second-kind
scaling. We further divide our discussions into two regimes, (1) a study of the
resonance region where $\Delta$'s and $N^*$'s are expected to play an
important role along with non-resonant meson production, and (2) a
study of the very limited data available in the Deep Inelastic
Scattering (DIS) region. Upon seeing that the second-kind 
scaling behavior is only moderately violated in these regimes, and
motivated by the type of analysis performed in studying the EMC
effect, we also define an appropriate ratio involving the $f_T$'s for a
pair of nuclei. This provides a convenient measure of the extent to
which scaling of the second kind is or is not respected. Indeed, as we
shall see, the ratio is very close to unity from the most negative
values of the scaling variable ({\it i.e.,} far below the quasielastic
peak) up through the resonance region. Only in the DIS region does the
ratio differ from unity by as much as 25\%; for most of the region the
results lie typically within about 10\% of unity.

The paper is organized as follows: in the next section we briefly
summarize the basic scaling formalism, drawing on our previous 
work \cite{don99,don99a} where more detailed discussions can be found.
In Sec.~\ref{sec:determ} the updated determinations of $k_F$ and $E_{shift}$ are
discussed, in Sec.~\ref{sec:trans} the transverse scaling functions $f_T$ are
presented and in Sec.~\ref{sec:ratios} the ratios involving pairs of nuclei are
introduced. For the last, the discussions are focused on two kinematic
regimes, the resonance region in Sec.~\ref{subsec:reson} and the DIS region in
Sec.~\ref{subsec:Dis}. In the former additional modeling is presented to help in
understanding why the ratios differ from unity even by the small
amount they do. Finally, in Sec.~\ref{sec:concl} we summarize our observations
and conclusions.

\section{Basic Scaling Formalism}
\label{sec:form}

We begin by summarizing some of the essential expressions used 
in previous studies of 
scaling and superscaling in the quasielastic region. First, 
using the nucleon mass $m_N$ 
as a scale, it proves useful to introduce
dimensionless variables to replace the 3-momentum transfer {\bf q} 
and energy transfer $\omega$, namely
\bea
\lambda &\equiv& \difrac{\omega}{2 m_N}   
\label{eq:lambda}\\
\kappavec  &\equiv& \difrac{{\bf q}}{2 m_N} 
\label{eq:kappa}.
\eea
The (unpolarized) inclusive electron scattering cross section 
then depends only on 
$\kappa=|\kappavec|$, $\lambda$ and the electron scattering 
angle $\theta_e$. The 
dimensionless 4-momentum transfer squared is given by
\be
\tau\equiv |Q^2|/4 m_N^2 = \kappa^2 - \lambda^2
\label{eq:tau},
\ee
and $Q^2=\omega^2 - q^2 < 0$ in the conventions used here.

In past work it became apparent that a convenient 
dimensionless scale in quasielastic 
electron scattering from nuclei is provided by the ratio 
of a characteristic nuclear 
momentum ${\bar k}=\sqrt{<k^2>}$ to the nucleon mass $m_N$. For example, in the 
Relativistic Fermi Gas (RFG) model the characteristic momentum is the Fermi 
momentum $k_F$ and the dimensionless scale is given by $\eta_F$, where
\be
\eta_F  \equiv \difrac{k_F}{m_N}\ll 1,        
\label{eq:etaF}
\ee
where typically the Fermi momenta range from as small as 
55 MeV/c for deuterium, 200 
MeV/c for $^4$He, to as large as about 250 MeV/c for very heavy nuclei, and as a 
consequence the strong inequality above holds. A corresponding 
dimensionless energy scale is also useful,
\be
\varepsilon_F \equiv \sqrt{1+\eta_F^2 } = 1 + \difrac{1}{2}\eta_F^2 + \cdots ,
\label{eq:varepsilon} 
\ee
where then
\be
\xi_F   \equiv \varepsilon_F -1   = \difrac{1}{2}\eta_F^2 + \cdots .
\label{eq:xiF}
\ee 
Naturally the RFG is only a first approximation to the nuclear 
dynamics involved in the 
quasielastic region and thus the Fermi momenta actually 
employed --- and these are 
obtained by fitting data as discussed below and in \cite{don99,don99a} --- 
should really be regarded as effective parameters in the 
problem. Clearly across the 
periodic table the densities of nuclei change and this is 
reflected in the fact that the 
characteristic momentum scale (here $k_F$) should also vary, 
roughly such that the density is proportional to $k_F^3$. As a
consequence the width of the quasielastic response goes as $k_F$.

In past studies of the region at and below the quasielastic 
peak it has proven to be very useful to introduce scaling 
variables (see, for example, \cite{Day90}). In the most familiar 
approach the $y$-scaling variable is employed and one finds 
that at high momentum transfers the experimental 
inclusive cross sections divided by an appropriate single-nucleon 
cross section scale, {\it i.e.,}
become functions only of $y$ and not independently of the energy
or momentum transfer. 
Such behavior is called 
{\em scaling of the first kind}. Alternatively, again using the 
RFG for guidance, a 
dimensionless scaling variable $\psi$ emerges naturally 
\cite{alb88,bar98}:
\be
\psi \equiv \difrac{1}{\sqrt{\xi_F}}
\difrac
{\lambda-\tau}
{ 
\sqrt{ (1+\lambda)\tau + \kappa \sqrt{ \tau(\tau+1) } } 
} .
\label{eq:psi}
\ee
At the na\"{\i}ve quasielastic peak where $\lambda=\tau$ (which corresponds to 
$\omega=|Q^2|/2 m_N$) one has $\psi=0$ and finds that the RFG 
response region is mapped into 
the range $-1\le \psi \le +1$. While appearing at first sight to be 
quite different, in fact the 
variables $y$ and $\psi$ are closely related. One can 
write \cite{don99a}
\be
\psi = \frac{y(E_s = 0)}{k_F} \left[ 1 + {\cal O}(\eta_F\psi,
y/M_{A-1}^0) \right] ,
\label{eq:psiappx}
\ee
where $E_s$ is the separation energy, the difference between 
the sum of the nucleon plus 
ground-state daughter masses and the target ground-state mass. 
Up to the choice made in 
making the scaling variable dimensionless by dividing by $k_F$ (see below), when 
$E_s$ is set to zero the two variables differ only at 
order $\eta_F$ and order $[M_{A-1}^0]^{-1}$, and accordingly when 
scaling of the first kind occurs in one it is bound to 
occur in the other, as long as corrections to the leading order
expressions are small. In effect, what the 
conventional $y$-scaling variable does that $\psi$ does not 
is to take into account the 
(small) shift in energy embodied in $E_s$. The simple
relationship begins to fail when large excursions are made
away from the QE condition $\psi=0$ --- in the present work the
full expressions are always used.

From such considerations one sees that an improved 
phenomenological dimensionless scaling variable 
can be employed in treatments of 
superscaling \cite{cen97}, namely one with an empirical 
shift $E_{shift}$. This was done in previous analyses
\cite{don99,don99a} introducing a dimensionless scaling
variable as above, 
\be
\psi^{\prime } \equiv \difrac{1}{\sqrt{\xi_F}}
\difrac{ \lambda^{\prime }-\tau^{\prime } }
{
\sqrt{ (1+\lambda^{\prime }) \tau^{\prime } 
      + \kappa \sqrt{ \tau^{\prime }(\tau^{\prime }+1) }
     }
} ,
\label{eq:psiprime}
\ee
where $\lambda_{shift} \equiv E_{shift}/2m_N$, 
$\lambda^{\prime } \equiv \lambda -\lambda_{shift}$
and $\tau^{\prime } \equiv \kappa^2 -\lambda^{\prime 2}$.

After this brief summary of the conventional choices of 
scaling variables, let us next turn 
to the scaling functions. We begin with the inclusive electron 
scattering cross section 
itself, which may be written in various forms (see also later):
\bea
d^2\sigma/d \Omega_e d\omega &=& 
{\Sigma}_L + {\Sigma}_T                    
\nonumber \\
&=&\sigma_M \left[ v_L R_L(\kappa,\lambda) 
+v_T  R_T(\kappa,\lambda) \right]           
\nonumber \\  
&=&\sigma_M \left[ W_2 +2 W_1 \tan^2 \theta_e/2)\right] ,  
\label{eq:sigmatot}
\eea
where the familiar electron kinematical factors in this 
Rosenbluth form are given by
\bea
v_L &=& \left[\difrac{\tau}{\kappa^2} \right]^2
\nonumber \\
v_T &=& \difrac{\tau}{2 \kappa^2} + \tan^2 \theta_e/2
\label{eq:vLT}
\eea
and the longitudinal (L) and transverse (T) response functions 
are related to $W_{1,2}$ 
via
\bea
R_T &=& 2 W_1
\nonumber\\
R_L &=& \left[ \difrac{\kappa^2}{\tau} \right]^2 W_2 
- \difrac{\kappa^2}{\tau} W_1 .
\label{eq:RLT-W12}
\eea

The strategy in discussing scaling in the quasielastic region 
is, to the extent that it is possible, to divide out the 
single-nucleon $eN$ elastic cross leaving only nuclear 
functions. As discussed in \cite{don99a,alb88,bar98}, this can be 
accomplished using the reduced response
\be 
F(\kappa,\psi) \equiv
\difrac
{ d^2 \sigma / d \Omega_e d\omega }
{
\sigma_M \left[ v_L G_L (\kappa,\lambda) 
 +v_T G_T (\kappa,\lambda) \right]
}                                                  
\label{eq:Ftot-RFG}
\ee
for the total cross section, or, when individual L 
and T contributions are being considered 
(as they are in part of the present work), using
\bea
F_L&\equiv&
\difrac
{R_L}{G_L (\kappa,\lambda)} 
\label{eq:FL-RFG}
\\
F_T&\equiv&
\difrac
{R_T}{G_T (\kappa,\lambda)} .
\label{eq:FT-RFG}
\eea
Here the functions $G_{L,T}$ are given by
\bea
G_L (\kappa,\lambda) 
&=& \frac{ (\kappa^2/\tau) 
[ {\tilde G}_E^2 + {\tilde W}_2 \Delta ] }
{2\kappa [1+\xi_F (1+\psi^2)/2]}   \label{eq:GLRFG} \\
G_T (\kappa,\lambda) 
&=& \frac{ 2\tau {\tilde G}_M^2 + {\tilde W}_2 \Delta }
{2\kappa [1+\xi_F (1+\psi^2)/2]}  
\label{eq:GTRFG}
\eea
which involve the function $\Delta$:
\bea
\Delta &=& \xi_F(1-\psi^2)
\left[
\difrac
{ \sqrt{\tau(1+\tau} } {\kappa} 
+\difrac{1}{3} \xi_F (1-\psi^2)
\difrac{\tau}{\kappa^2}
\right] 
\nonumber\\ 
&=& \difrac{1}{2}
(1-\psi^2)\eta_F^2 +{\cal O}[\eta_F^3] .
\label{eq:Delta} 
\eea
As usual one has
\bea
\tilde{G}_E^2 &\equiv& ZG_{Ep}^2 + NG_{En}^2 \nonumber \\
\tilde{G}_M^2 &\equiv& Z G_{Mp}^2 + NG_{Mn}^2 \nonumber \\
\tilde{W}_1 &=& \tau \tilde{G}_M^2 \nonumber \\
\tilde{W}_2 &=& \frac{1}{1+\tau} \left[\tilde{G}_E^2 
  + \tau \tilde{G}_M^2 \right] ,
\label{eq:tildes}
\eea
involving the proton and neutron Sachs 
form factors $G_{Ep,n}$ and $G_{Mp,n}$ 
weighted by the proton and neutron numbers $Z$ and $N$, 
respectively.

In the region of the QE peak where $|\psi|$ is small it is a 
good approximation to set $\Delta$ to zero and to expand the
functions $G_{L,T}$ in powers of $\eta_F^2$, retaining only the
lowest-order terms, namely to use
\bea
G_L (\kappa,\lambda) 
&=& \frac{\kappa}{2\tau} {\tilde G}_E^2 
+ {\cal O}[\eta_F^2] 
\label{eq:GLRFG-appx} 
\\
G_T (\kappa,\lambda) 
&=& \frac{\tau}{\kappa} {\tilde G}_M^2 + {\cal O}[\eta_F^2] .
\label{eq:GTRFG-appx}
\eea
While such leading-order expansions in $\eta_F^2$ are very good
when $\eta_F\times |\psi|$ is small, they become less so when
very large excursions away from the QE peak are made and accordingly
in the present work we have always used the full expressions in
Eqs.~(\ref{eq:GLRFG}--\ref{eq:GTRFG}).

As discussed in \cite{alb88,bar98} the RFG model then 
yields scaling of the first kind, namely, the $F$'s become functions only of 
$\psi$ (independent of $\kappa$, that is, of $q$); indeed, as 
discussed in \cite{don99,don99a}, so do the data when $\psi <0$. 
Moreover, as also discussed in \cite{don99,don99a,alb88}, the 
RFG model and the data both display scaling of the 
{\em second kind} in that the $F$'s 
can be made independent of the momentum scale in the 
problem --- that is, independent 
of $k_F$ to the order considered in the expansion in the 
small dimensionless parameter 
$\eta_F$. This is accomplished by defining
\bea
f&\equiv& k_F \times F
\label{eq:ftotal}
\\
f_{L,T}&\equiv&k_F \times F_{L,T}   ,    
\label{eq:fLT}
\eea
namely by making them dimensionless through multiplication by the factor $k_F$. 
Comparing Eq.~(\ref{eq:psiappx}) with the above, we see that the 
mapping of the $F$'s 
versus $y$ to the $f$'s versus $\psi$ (or $\psi^{\prime}$) 
is area-conserving: the 
dimensionless scaling variables contain a factor $k_F^{-1}$, 
while the scaling functions 
a factor $k_F^{+1}$. In the RFG model the $f$'s display 
scaling of both the first and 
second kinds, namely, the display {\em superscaling}.

\section{Determination of $k_F$ and $E_{shift}$}
\label{sec:determ}

In \cite{don99,don99a} the approach summarized in the previous
section was applied to an analysis of the usable data involving
inclusive electron scattering in the region of the quasielastic peak.
While medium-energy results were included, the main emphasis in that
study was placed on the high-energy results from SLAC and from TJNAF
\cite{Whitney74}--\cite{Blatchley86}.
And, being the first attempt to explore scaling of the second kind and
superscaling, we chose not to perform an extensive search 
to find the ``best'' choices of the two
parameters involved in the fit, namely, $k_F$ and $E_{shift}$, but 
instead selected values that were ``reasonable''. Now, given the success of
that previous study and the fact that scaling of the second kind
appears to be quite well obeyed in the scaling region ($\psi^\prime
<0$), we have stronger motivation to produce even better fits and to
assess the uncertainties in the fit parameters.

In the next section we shall place our focus on the transverse scaling
function $f_T$ and it would be desirable to adjust $k_F$ and
$E_{shift}$ for each nuclear species for this quantity. Unfortunately,
in the regime where $\psi^\prime\ll 0$, it has not been
possible to separate the longitudinal and transverse inclusive
responses and thus we are forced to make our fits to the total
$f$'s. Our hope is that the parameters obtained from fits to the total
are also appropriate for the individual $f_{L,T}$.

The total $f$'s are very sensitive functions of $k_F$ in the region
where $\psi^\prime\ll 0$ and yet it is possible to find values of
$k_F$ for which the data line up extremely well (see, for instance, 
the high-$q$/small-$\psi^\prime$ data shown below). The
$E_{shift}$ dependence is less critical than is that on
$k_F$; nevertheless, by examining the behavior near the quasielastic
peak it is clear that some shift is needed to move the response from
the na\"{\i}ve peak value of $\omega=|Q^2|/2 m_N$ ($\lambda=\tau$) to where
the data require it to be. In Table~1 we list the parameters obtained
from global fits to the data \cite{Whitney74}--\cite{Blatchley86}.

\vspace{0.3in}
\begin{center}
\begin{tabular}{|l||c|c|}
\hline
\multicolumn{3}{|c|}{Table 1: Adjusted Parameters}
\\ \hline \hline
\multicolumn{1}{|l||}{Nucleus} & $k_F$ (MeV/c) 
& $E_{shift}$ (MeV) \\ \hline \hline
Lithium    & 165  & 15 \\ \hline
Carbon     & 228  & 20 \\ \hline
Magnesium  & 230  & 25 \\ \hline
Aluminum   & 236  & 18 \\ \hline
Calcium    & 241  & 28 \\ \hline
Iron       & 241  & 23 \\ \hline
Nickel     & 245  & 30 \\ \hline
Tin        & 245  & 28 \\ \hline
Gold       & 245  & 25 \\ \hline
Lead       & 248  & 31 \\ \hline
\end{tabular}
\end{center}
\vspace{0.3in}

Clearly the energy shift does not vary too much. Presumably it
incorporates the separation energy $E_s$, the mean binding energy of
nucleons in the nucleus and some global aspects of final-state
interactions (for instance, RPA correlations which are known to shift
the response slightly). While attempts are being made to account for
the values found here, it is not a simple problem to address for the
relatively high-energy conditions of most of the current study where
relativistic effects are known to be very important, and thus here
we limit ourselves to the present phenomenological discussion.

The values of $k_F$ found vary monotonically once ``typical'' nuclei
--- say, beyond carbon --- are reached. It should be understood that
the values given here are relative and not absolute: if all values are
scaled by a common factor, then equally good scaling
of the second kind is obtained. The
value of 228 MeV/c for carbon is typical of other studies and so we
have used this to normalize the rest. The present fits are done
emphasizing the large negative $\psi^\prime$ region where
``contamination'' from pion production, 2p-2h MEC effects, resonances
and DIS are thought to be small and where the basic underlying nuclear
spectral function is presumably revealed most clearly, in contrast to
some previous attempts to determine $k_F$ using the entire response
region. Interestingly the values found here for heavy nuclei are 
somewhat smaller than those generally chosen --- lead, for
example, is sometimes assumed to have $k_F\cong$ 265 MeV/c. We
believe that the present values are more reliable determinations of
the effective $k_F$'s. Note also the curious value for lithium,
curious because the $k_F$ used for $^4$He is 200 MeV/c (see
\cite{don99,don99a}). However, this is easily explained if one assumes
that $^6$Li is essentially a deuteron (with $k_F=$ 55 MeV/c) 
plus an alpha-particle. Taking the weighted mean $[\{4\times (200)^2 + 
2\times (55)^2\}/6]^{1/2}$ gives 166 MeV/c, which is very close to the
fit value of 165 MeV/c.

\begin{figure}[htb]
\centerline{\mbox{\epsfysize=14cm \epsffile{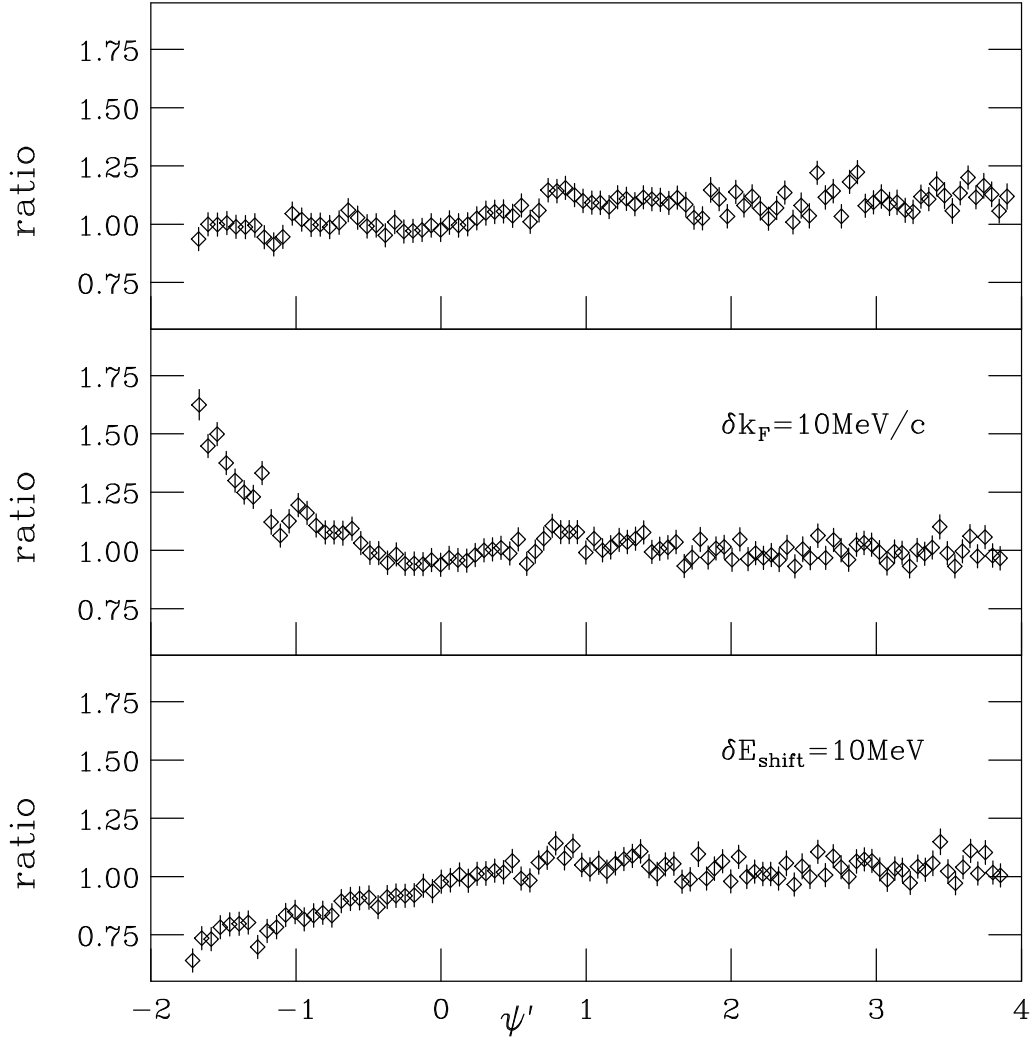}}}
\begin{center} 
\parbox{15cm}
{ \caption{ \label{fig:fratio}  The ratio of $f_{Au}$ over $f_{C}$ for
    energy 3.6 GeV and scattering angle 16 degrees. The top panel
    shows the ratio using $[k_F]_{Au}=245$ MeV/c,
    $[E_{shift}]_{Au}=25$ MeV and $[k_F]_{C}=228$ MeV/c,
    $[E_{shift}]_{C}=20$ MeV, the ``best fit'' values. In the middle
    panel $[k_F]_{Au}$ has been increased by 10 MeV/c, while in the
    bottom panel $[E_{shift}]_{Au}$ has been increased by 10 MeV.
}}   
\end{center}
\end{figure}

To get some feeling for the sensitivity of the fits in
Fig.~\ref{fig:fratio} we show the ratio $f_{Au}/f_{C}$ for data from
SLAC taken at $\theta_e=$ 16 degrees and incident electron energy 
$E_e=$ 3.6 GeV --- for more
discussion of the existing data used \cite{Whitney74}--\cite{Blatchley86}
see \cite{don99,don99a}. The top panel in the figure
shows that in the region $\psi^\prime <0$ the data themselves scatter 
at roughly the 10\% level, {\it i.e.}, scaling of the second kind for these
kinematics is satisfied to roughly the 10\% level, which is clearly
better than scaling of the first kind (see \cite{don99,don99a} and
figures given below). At positive $\psi^\prime$ the ratio moves above
unity and constitutes the focus of the discussions later in the
present work. In the middle panel the $k_F$ of gold has been increased
by 10 MeV/c: clearly the fit is very poor in the
negative $\psi^\prime$ region, indicating that the values of $k_F$
given in Table~1 are rather finely determined, namely, to only a few
MeV/c. In the bottom panel in the figure, the energy shift used for gold is
increased by 10 MeV and again the fit is much worse
than the ``best-fit'' value. So the values of $E_{shift}$ given in the
table are good to perhaps a few MeV. Finally, note that, in the large
positive $\psi^\prime$ region which will be discussed in depth below,
the sensitivity to variations in either parameter is much weaker. This
stems from the rapid all-off of the responses below $\psi^\prime=0$,
in contrast to the relatively flat response when $\psi^\prime>0$.

\section{Transverse Scaling Functions}
\label{sec:trans}

\begin{figure}[htb]
\centerline{\mbox{\epsfysize=11cm \epsffile{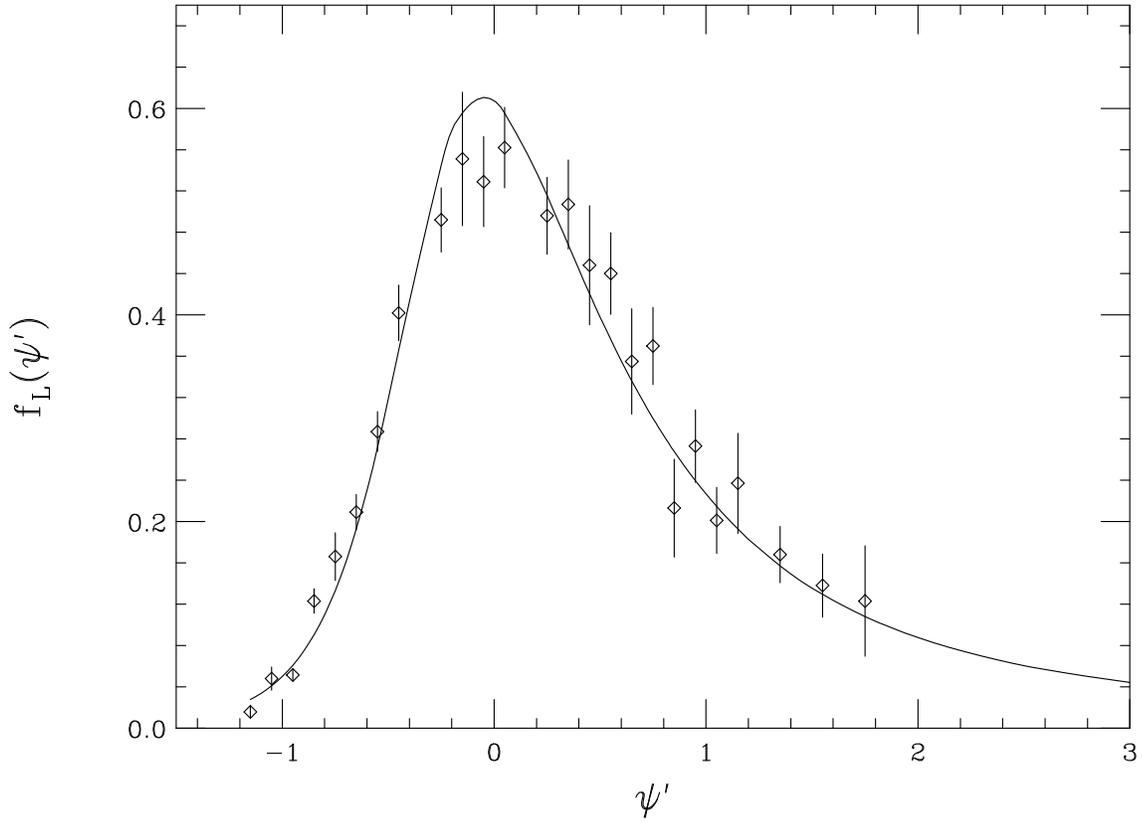}}}
\begin{center} 
\parbox{15cm}
{ \caption{ \label{fig:flong}  Averaged $f_L(\psi ')$ together 
with a convenient parameterization of the results.
}}   
\end{center}
\end{figure}

From our previous analysis \cite{don99a} we have seen indications that
the longitudinal scaling function $f_L$ exhibits {\em superscaling}
behavior, that is, it not only displays scaling behavior of the second
kind (as does the total $f$ discussed above), but it has scaling
behavior of the first kind. Of course, the regime in which this
superscaling has been verified is relatively limited, given the difficulty of
separating the longitudinal and transverse response functions. 
In practice, using the analysis in \cite{jour96} we have made a 
fit to the combined set of $f_L$-values for the higher momentum 
transfers where scaling of the first kind is seen to occur. The
results are shown in Fig.~\ref{fig:flong}.

To make any further progress on the problem, it is necessary to make an
assumption and we do so now: we assume that the so-determined
longitudinal scaling function shown in Fig.~\ref{fig:flong}
is universal ({\it i.e.}, superscaling
works). Given this universal $f_L$ we can immediately reconstruct the
longitudinal cross section for any kinematical condition using the
expressions given in Sec.~\ref{sec:form},
\be
\Sigma_L=\frac{1}{k_F}f_L \sigma_M v_L G_L ,
\label{eq:SigmaL}
\ee
from this isolate the transverse part of the cross section,
\be
\Sigma_T=\frac{d^2\sigma}{d\Omega_e d\omega} - \Sigma_L
\label{eq:SigmaT}
\ee
and so obtain the transverse scaling function,
\be
f_T (\psi^\prime)= \frac{\Sigma_T}{\sigma_M v_T G_T} .
\label{eq:ftfromf}
\ee

\begin{figure}[htb]
\centerline{\mbox{\epsfysize=18cm \epsffile{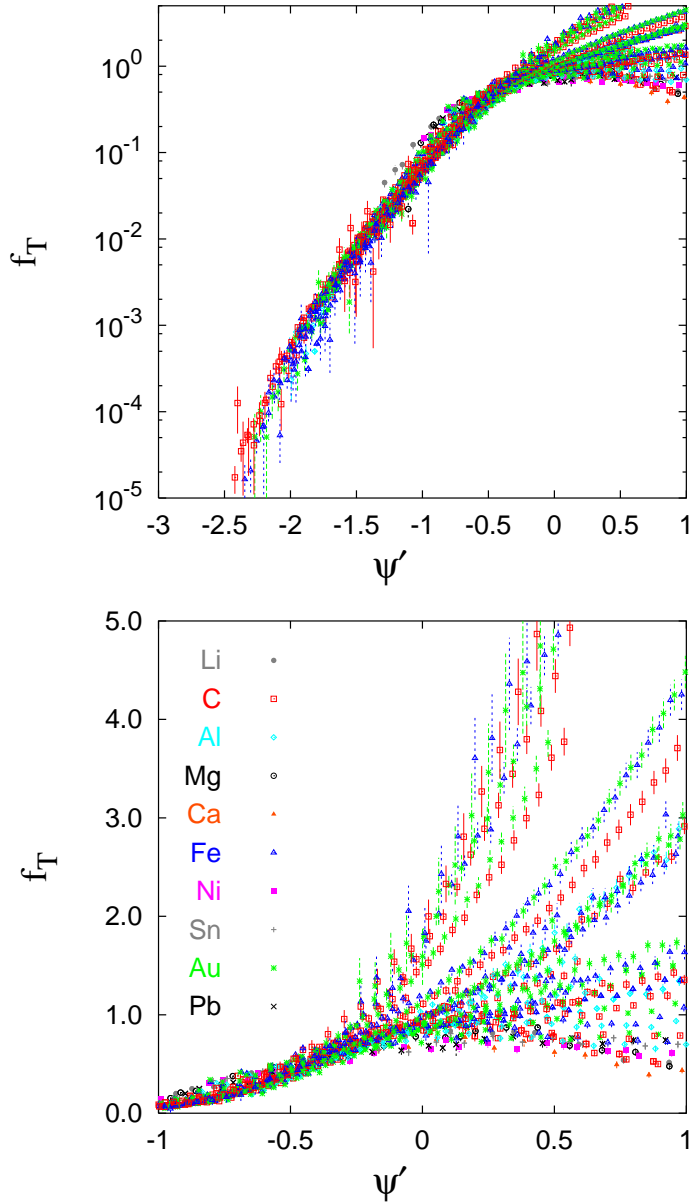}}}
\begin{center} 
\parbox{15cm}
{ \caption{ \label{fig:ftall} (Color) Transverse scaling function $f_T$ for a
    wide range of nuclei and for kinematics ranging from medium
    energies (500 MeV and 60 degrees) to high energies (up to 4.045
    GeV and 74 degrees). The longitudinal response has been removed
    using the superscaling assumption discussed in the text. The fit
    parameters are listed in Table~1.
}}   
\end{center}
\end{figure}

Using this procedure we arrive at the transverse scaling function
$f_T$. In Fig.~\ref{fig:ftall} we show the results obtained for all
kinematics from medium-energy measurements at 500 MeV and 60 degrees
($q\cong 0.4$ GeV/c) to results from both SLAC and TJNAF ranging up to
$q\cong 4$ GeV/c. For $\psi^\prime < -0.3$ we see a reasonable
convergence of the results to a band, although the width of the band
is not negligible, reflecting (at least) breaking of scaling of the first
kind. Since the span of momentum transfers is so large in the results
shown in the figure, we are emphasizing the lack of 1$^{st}$-kind
scaling, and focusing on a smaller range of $q$ produces less spread,
as discussed below. Note also that the region above $\psi^\prime\cong
-0.3$ contains a very large spread, that is, a very large degree of
scale-breaking. A motivation of the present work is to begin to get
some insight into the nature of this behavior.

\begin{figure}[htb]
\centerline{\mbox{\epsfysize=18cm \epsffile{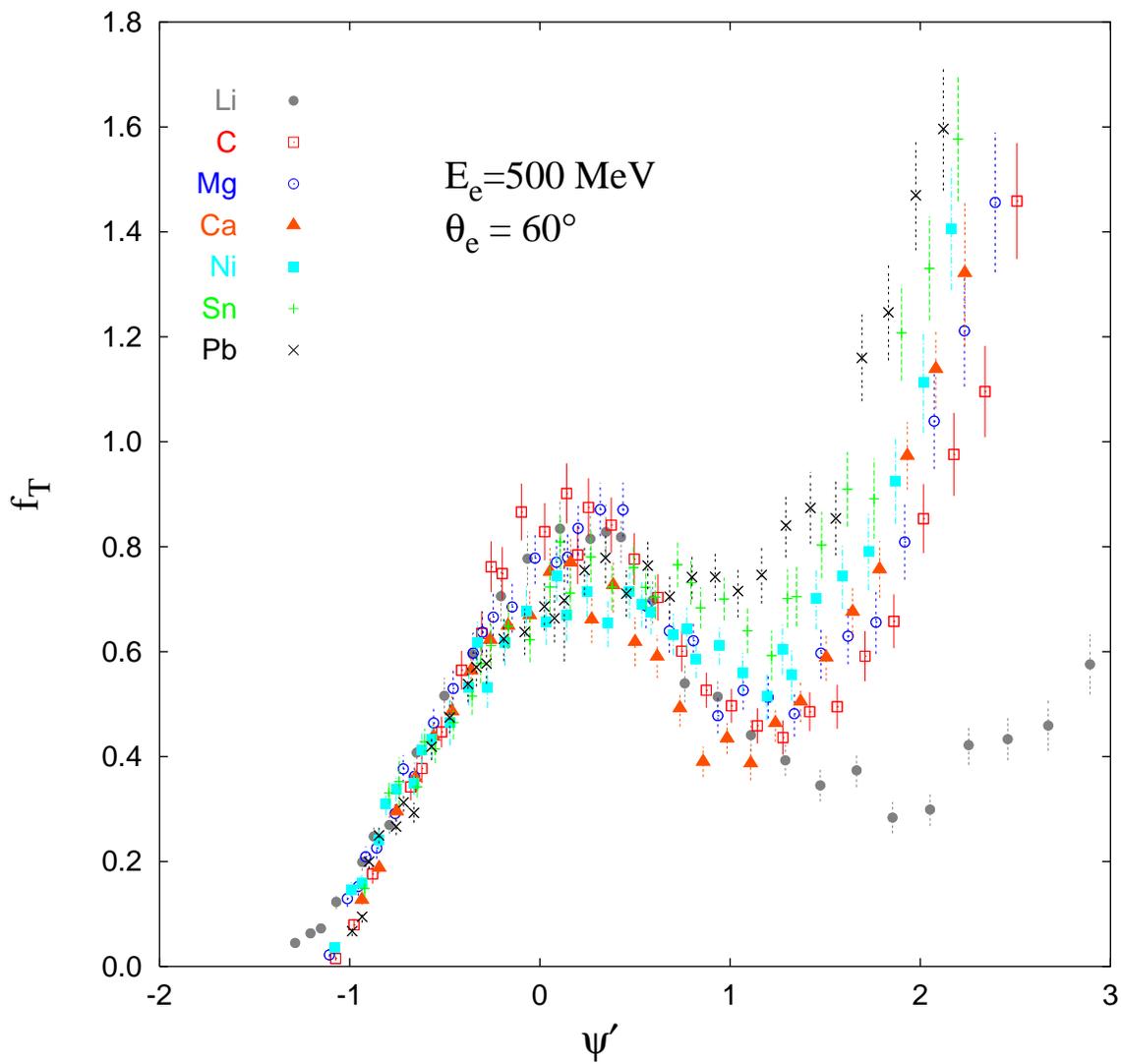}}}
\begin{center} 
\parbox{15cm}
{ \caption{ \label{fig:ftmed}  (Color) As in 
            Fig.~\protect\ref{fig:ftall}, 
            but only for medium energy data (500 MeV and 60 degrees).
}}   
\end{center}
\end{figure}

In Fig.~\ref{fig:ftmed} we show only the medium-energy results for
$f_T$. Here, at energy 500 MeV and scattering angle 60 degrees, the momentum
transfer varies from about 490 MeV/c at $\psi^\prime = -1$ down to
about 430 MeV/c at the largest values of $\psi^\prime$. We see on the
one hand, that now the band in the negative-$\psi^\prime$ region is fairly
tight, an indication that the breaking seen in Fig.~\ref{fig:ftall} is
indeed mainly due to 1$^{st}$-kind scale breaking and not to 2$^{nd}$-kind
breaking. On the other hand, the behavior at positive-$\psi^\prime$
shows that there one also has breaking of 2$^{nd}$-kind scaling
behavior. Clearly as one proceeds from light nuclei with low $k_F$ to
heavy nuclei with large $k_F$ the trend is to increasingly large
values of $f_T$ in this region. This indicates that the mechanisms
that produce the scale-breaking must go as some positive power of
$k_F$. Indeed, in recent work \cite{ama01,ama01a,car01} such breaking of
both 1$^{st}$- and 2$^{nd}$-kinds due to MEC and correlation effects in
the 1p-1h sector has been investigated in detail 
and work is in progress to arrive
at a relativistic extension of older work \cite{van81} in which
scale-breaking in the 2p-2h sector was also identified.

\begin{figure}[htb]
\centerline{\mbox{\epsfysize=18cm \epsffile{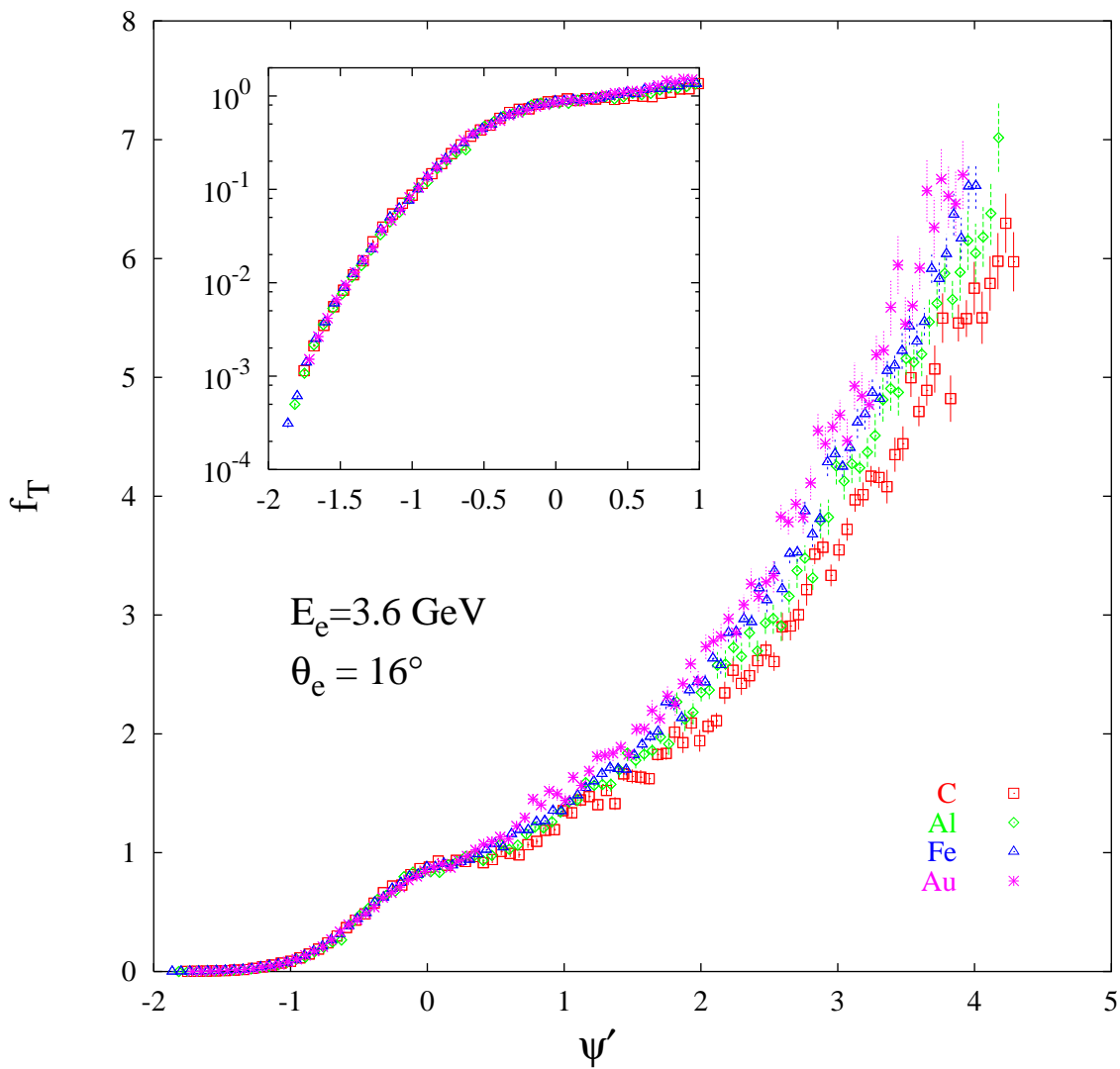}}}
\begin{center} 
\parbox{15cm}
{ \caption{ \label{fig:ftslac} (Color)  
   As in Fig.~\protect\ref{fig:ftall}, but only
   for data at 3.6 GeV and 16 degrees.
}}   
\end{center}
\end{figure}

Next, in Fig.~\ref{fig:ftslac} we show $f_T$ for SLAC data at 3.6 GeV
and 16 degrees scattering angle. At the lowest values of $\psi^\prime$
the momentum transfer is roughly 990 MeV/c while at $\psi^\prime\cong
4$ it has risen to about 1.7 GeV/c. As the inset on a semi-log scale
clearly shows, the quality of second-kind scaling behavior at higher
$q$-values is excellent in the negative-$\psi^\prime$ region. At
positive $\psi^\prime$ 2$^{nd}$-kind scaling is not perfect, although it
is only modestly broken. The lower the value of $k_F$ the smaller is $f_T$,
indicating again that the scale-breaking mechanisms go as some
positive power of $k_F$, as expected.

\begin{figure}[htb]
\centerline{\mbox{\epsfysize=18cm \epsffile{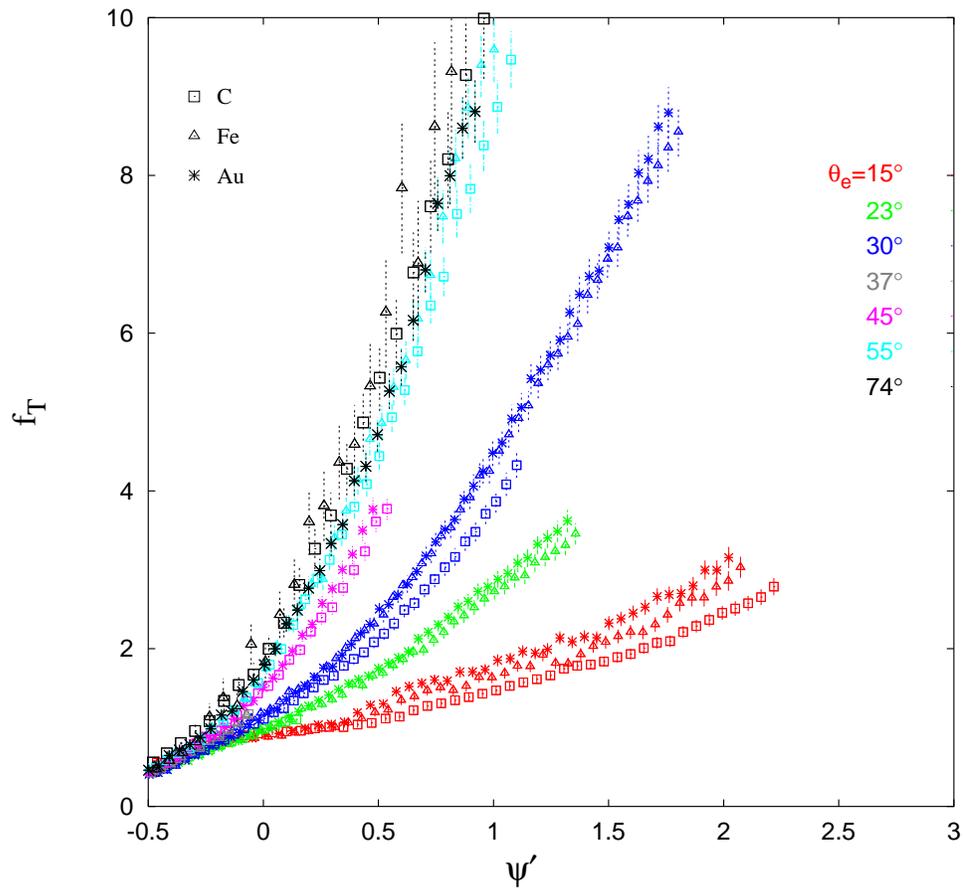}}}
\begin{center} 
\parbox{15cm}
{ \caption{ \label{fig:ftcebaf} (Color)  As in 
            Fig.~\protect\ref{fig:ftall}, but  for data at 4.045 GeV 
            and angles ranging from 15 to 74 degrees.
}}   
\end{center}
\end{figure}

Finally, in Fig.~\ref{fig:ftcebaf} we show several sets of data taken
at 4 GeV energy and various scattering angles. In this single figure one is
able to assess scale-breaking of both kinds. Each fairly
tightly-grouped set of data displays the extent of 2$^{nd}$-kind
scale-breaking and clearly the results are very similar to those
already seen in the previous figure. On the other hand, the various
scattering angles yield correspondingly different values for the 
momentum transfer, ranging, for instance at $\psi^\prime=+1$ from
about 1.2 GeV/c at 15 degrees to 3.9 GeV/c at 74 degrees. For each
different choice of kinematics a different grouped set of results is
obtained, indicating again the extent to which scaling of the
1$^{st}$-kind is broken.

In summary, we deduce from these results for $f_T$ that in the region
below $\psi^\prime\cong 0$ scaling of the 2$^{nd}$-kind is excellent,
while scaling of the 1$^{st}$ is violated, although not too badly. In
contrast, for $\psi^\prime >0$ scaling of the 2$^{nd}$-kind is also
broken to some extent, whereas scaling of the 1$^{st}$-kind is very
badly broken. Accordingly, let us next turn to a closer examination of
the region above the quasielastic peak to see if further insight can
be obtained on how the moderate scale-breaking of the 2$^{nd}$-kind
arises.

\section{Ratios of Transverse Scaling Functions}
\label{sec:ratios}

\begin{figure}[htb]
\centerline{\mbox{\epsfysize=18cm \epsffile{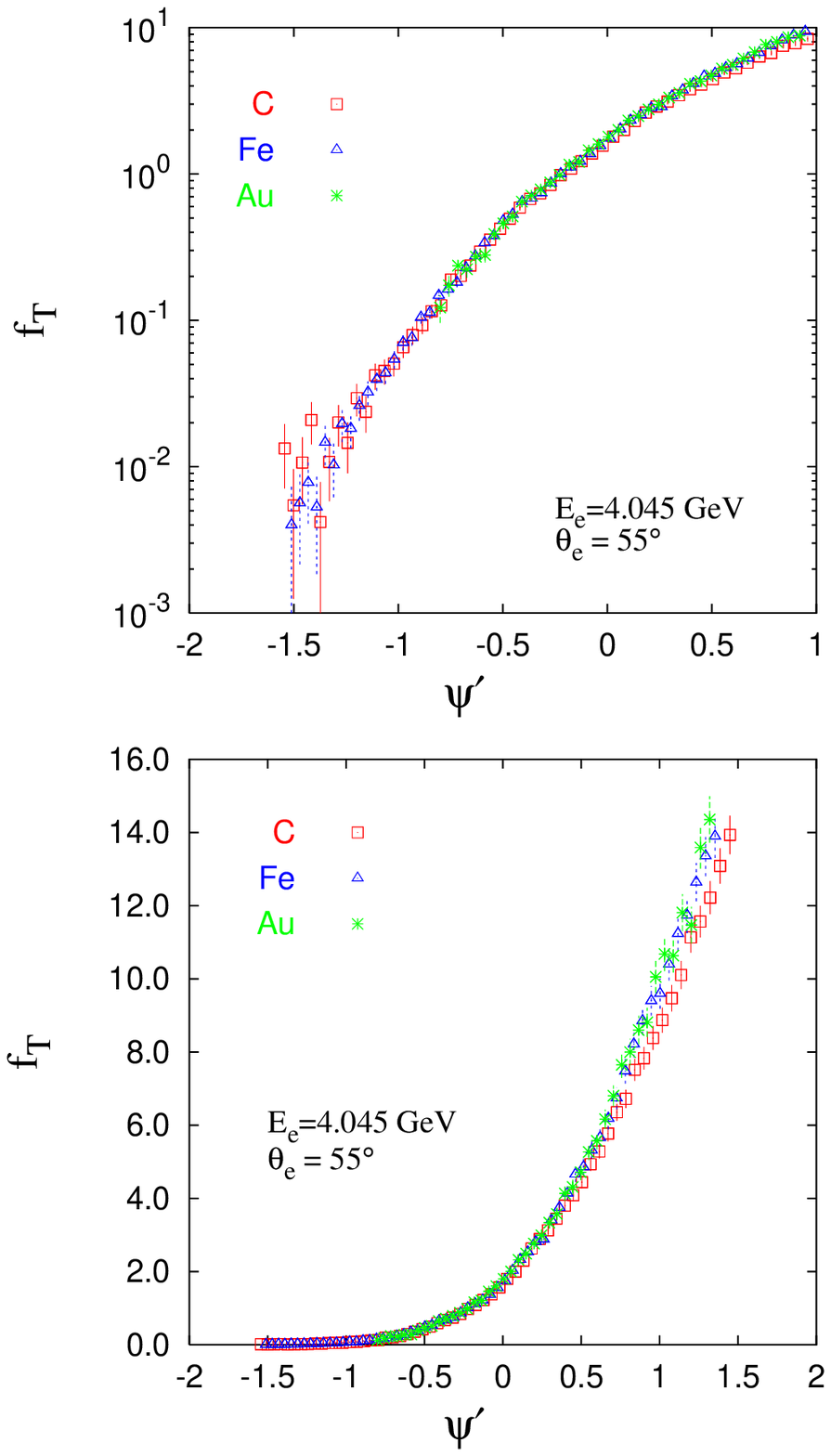}}}
\begin{center} 
\parbox{15cm}
{ \caption{ \label{fig:ftcebaf55} (Color)  As in 
  Fig.~\protect\ref{fig:ftcebaf}, but only
    for data at 4.045 GeV and 55 degrees.
}}   
\end{center}
\end{figure}

The results given in the previous section clearly show that several
different regions are involved when studying the first- and
second-kind scaling behaviors of $f_T$. In the region below the
quasielastic peak ($\psi^\prime < 0$) at high-$q$ the first-kind
scaling is reasonably good --- this is what is usually called simply
$y$-scaling --- and the second-kind scaling is excellent. Another
example is shown in Fig.~\ref{fig:ftcebaf55} containing TJNAF data at
4 GeV and 55 degrees ($q\cong 3.4$--3.5 GeV/c). Clearly one could
present the results as a {\em ratio}, that is, as 
$\left[ f_T(\psi^{\prime})\right]_2/\left[
  f_T(\psi^{\prime})\right]_1$, where 1 and 2 denote two different
nuclei. For $\psi^\prime < 0$ one would
obtain unity with very small uncertainties at all but the lowest 
values of the scaling variable, while for $\psi^\prime > 0$ the ratios
deviate from unity.  In particular, if 1 $\leftrightarrow$ light nucleus
and 2 $\leftrightarrow$ heavy nucleus, then the ratio rises above unity
in the region above the quasielastic peak. For instance, in 
Fig.~\ref{fig:ftcebaf55} at $\psi^\prime = +1$ the Au/C ratio is about
1.2. Thus, even at such high values of the momentum transfer the
ratio in the entire region from very large negative $\psi^\prime$ to very large
positive $\psi^\prime$ falls within roughly 20\% of unity.

Let us now focus on the region above the quasielastic peak and see if
more light can be shed on the remaining (modest, {\it i.e.,}
$\approx$20\%) second-kind scale breaking. To begin with, note that
the scaling functions $f$, $f_L$ and $f_T$ have been constructed with
the quasielastic responses in mind: the denominator in
Eq.~(\ref{eq:Ftot-RFG}) contains the elastic eN form factors weighted
by $Z$ and $N$, the proton and neutron numbers,
respectively. In the region below the quasielastic peak we expect the
dominant contributions to the response functions to arise from the
tails of the QE response and accordingly removing such a weighted
elastic eN cross section makes sense. However, in the region above the
quasielastic peak we do not expect things to be so simple. Certainly
resonance excitation and pion production play an important role and,
at asymptotically high values of momentum transfer, one expects DIS to
take over. Because of this one expects the longitudinal and transverse
responses to have quite different characters --- for instance, the
first important new contribution beyond QE scattering in entering the
$\psi^\prime>0$ region is expected to come from $N\to\Delta$
excitations on nucleons moving in the nucleus, and this is (1)
essentially transverse and (2) isovector. For this reason we have
extracted the transverse response using the superscaling assumption
for the longitudinal contributions. Then, focusing on the transverse
results, we can attempt variations on the theme that motivated
$y$-scaling in the first place. Specifically, instead of weighting
protons and neutrons with $G_{Mp}^2$ and $G_{Mn}^2$, respectively, we
can make a scaling function that varies only with total mass number
$A=Z+N$ and not $Z$ and $N$ individually --- this is more in the
spirit of what is done for the ratio considered when studying the EMC
effect. For instance, if an isovector excitation such as $N\to\Delta$ 
dominates in the response, then equal weighting of protons and
neutrons should be better that the weighting provided using the
elastic magnetic form factors of the nucleons. Accordingly let us consider
\be
{\bar f}_T\equiv \frac{k_{F}\Sigma _{T}}{A\sigma _{M}v_{T}
\left( \frac{\tau }{\kappa }\right) G_{D}^{2}} ,
\label{eq:xi}
\ee
where $G_D$ is the dipole form factor 
(a compromise between the $Q^2$ dependences of $G_{Mp}$ and $G_{Mn}$
--- its form is not important here as it will cancel in constructing
the ratio defined below). This 
provides an alternative for $f_T$ defined above in 
Eqs.~(\ref{eq:FT-RFG},\ref{eq:fLT}). The two differ roughly by a factor
\bea
\overline{\mu}^{2} &\equiv &
\frac{Z G_{Mp}^{2}+N G_{Mn}^{2}}{A G_{D}^{2}}        
\nonumber \\
&\cong &\frac{Z }{A}\mu _{p}^{2}+\frac{N}{A}\mu _{n}^{2} \nonumber \\
&\cong & 5.73 \left[ 1 - 0.36 (N-Z)/A  \right] ,
\label{eq:muappx}
\eea
which is a relatively slowly-varying function of (Z, N). 
Such an approximation will allow 
us to interconnect the quasielastic and DIS analyses in an 
approximate way (see below).

With these definitions, let us next consider two different 
nuclei have charge and neutron 
numbers (Z$_i$, N$_i$) with $i=1,$ 2. Labelling as above 
with nuclear species numbers 
1 and 2, we consider the ratio
\be
\rho _{12}\equiv \difrac{\left[ {\bar f}_T(\psi^{\prime})\right]_2}
       {\left[ {\bar f}_T(\psi^{\prime})\right]_1} .
\label{eq:rho}
\ee
With $\overline{\mu}^{2}_{12} \equiv
\overline{\mu_1}^2/\overline{\mu_2}^2$ we then have that
\be
\difrac{\left[ f_T(\psi^{\prime})\right]_2}
       {\left[ f_T(\psi^{\prime})\right]_1}
\cong \overline{\mu}^{2}_{12} \times \rho _{12} \cong \rho _{12} ,
\label{eq:ratft}
\ee
where the last approximation holds to the degree that 
\be
\overline{\mu}^{2}_{12} 
\cong  \frac{1 - 0.36 (N_1-Z_1)/A_1}{1 - 0.36 (N_2-Z_2)/A_2}
\label{eq:mu12}
\ee
is nearly unity across the periodic table. For instance, in
going from a nucleus such as carbon ($N_1=Z_1$) to gold
($N_2=118$, $Z_2=79$) one has $\overline{\mu}^{2}_{12} 
\cong 1.08$. Thus the ratio of the $f$'s will
be about 8\% larger than the quantity $\rho_{12}$ 
for gold/carbon and even closer to unity for ratios involving nuclei
closer together in the periodic table.

From this simple analysis we can see that some of the rise discussed
above for the ratio 
$\left[ f_T(\psi^{\prime})\right]_2/\left[
  f_T(\psi^{\prime})\right]_1$ in the region $\psi^\prime>0$ occurs
because of the different weighting of protons and neutrons in
regions dominated by processes other than quasielastic
scattering. Two examples of the ratio $\rho_{12}$ are shown in 
Fig.~\ref{fig:r12slac}. These have been obtained by extracting the 
transverse contributions from the experimental cross sections for each
nucleus using the superscaling assumption to remove the longitudinal 
contributions, as discussed in the previous section, calculating 
${\bar f}_T(\psi^{\prime})$ for each nucleus using Eq.~(\ref{eq:xi}),
fitting the results and finally using Eq.~(\ref{eq:rho}) for specific pairs
of nuclei. The upper panel in the figure corresponds to momentum transfers of
about 1.2 GeV/c (near $\psi^\prime=1$) to 1.7 GeV/c (near
$\psi^\prime=4$), whereas the lower panel ranges from about 2.1 GeV/c
(near $\psi^\prime=1$) to 2.4 GeV/c (near $\psi^\prime=2.5$).
Thus, as the momentum transfer grows the ratio seems to stabilize at
roughly 5--10\% above unity (no error bands are shown in the figure,
but typically the uncertainties are $\pm$2\%). 
The analysis in terms of $\rho_{12}$ accounts for some of the
2$^{nd}$-kind scale breaking in the region above the quasielastic peak;
however, it still does not explain the full effect, but
only perhaps half of the 20\% discussed above. To get more insight we
need to invoke a model of the transverse response in the region
$\psi^\prime>0$. We proceed in the following two subsections by
examining the resonance region and DIS separately.

\subsection{In the Resonance Region}
\label{subsec:reson}
\begin{figure}[htb]
\centerline{\mbox{\epsfysize=18cm \epsffile{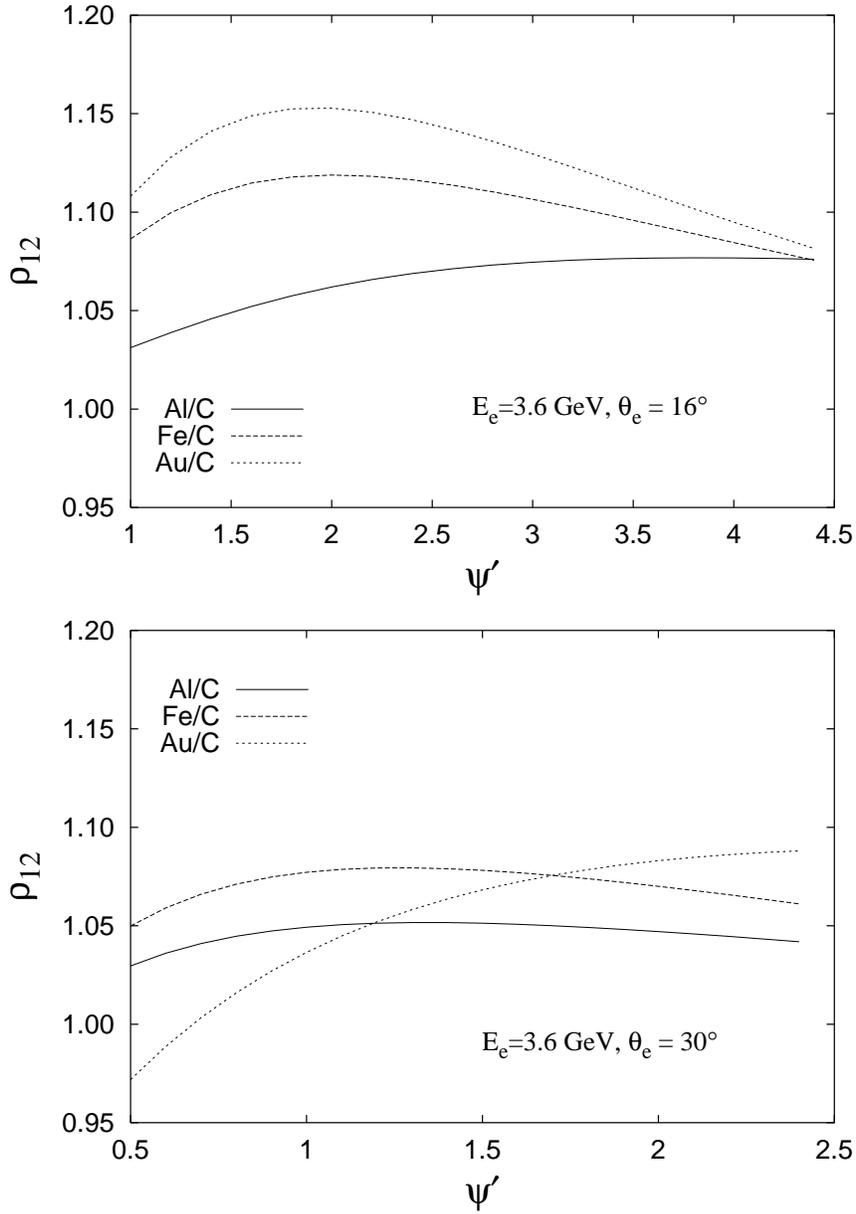}}}
\begin{center} 
\parbox{15cm}
{ \caption{ \label{fig:r12slac}  The ratio $\rho_{12}$ defined in the
    text for two kinematic conditions, 3.6 GeV with 16 and 30 degrees.
}}   
\end{center}
\end{figure}

Let us begin by focusing on the region in which we expect the
excitation of resonances and meson production to become relevant, in
addition to the tail of the quasielastic response. In
Fig.~\ref{fig:r12slac} this corresponds to $\psi^\prime\sim 1$--2
(upper panel) and somewhat smaller (lower panel); see below where the
scaling variable $\psi_*$ is introduced for a more quantitative
measure of where a given baryon resonance has its peak.
In \cite{ama99} an extension of the RFG model was discussed. Instead
of building the model from elastic scatterings of electrons from
nucleons in this work the inelastic $N\to\Delta$ transition was
considered and again a relativistic Fermi gas model constructed. Of
course the ideas can be extended even further to inelastic excitation
of any baryon resonance, that is, with any mass $m_*$. Following
\cite{ama99} the RFG hadronic tensor for the production of a stable
resonance of this mass is given by
\bea
W_{\mu\nu} &=& \difrac{3\pi^2 {\cal{N}} m_N^2}{k_F^3}
\int \difrac{d \pvec}{E(\pvec)E_*(\pvec +\qvec)} \theta(k_F-p)
f_{\mu\nu}(\pvec,\pvec+\qvec)                      \nonumber\\
&& \qquad\qquad\qquad 
\times \delta\left[E_*(\pvec +\qvec) -E(\pvec) -\omega\right] \;,
\label{eq:Wmunu-Delta}
\eea
where $\cal{N}$ is the number of protons or neutrons (the total
nuclear response should be $Z$ times the result for target protons
plus $N$ times the result for target neutrons). As usual 
$E(\pvec)= \sqrt{p^2 +m_N^2}$ and now also
$E_*(\pvec +\qvec)= \sqrt{(\pvec +\qvec)^2 +m_*^2}$. 
Here $f_{\mu\nu}$ is the single-nucleon {\em inelastic} hadronic tensor:
\bea
f_{\mu\nu}(\pvec,\pvec+\qvec) &=&
- w_1(\tau)\left[g_{\mu\nu} - \frac{Q_\mu Q_\nu}{Q^2}\right] 
\nonumber \\
&& +w_2(\tau)\frac{1}{m_N^2}\left[P_\mu - \frac{P\cdot Q}{Q^2}Q_\mu\right]
\left[P_\nu - \frac{P\cdot Q}{Q^2}Q_\nu\right] ,
\label{eq:fmunu-Delta}
\eea
where $w_{1,2}$ are the analogs of $W_{1,2}$ for elastic scattering
--- the $w$'s for the $N\to\Delta$ case are discussed in \cite{ama99};
however, for the present purposes it is sufficient simply to know that
they are functions only of the 4-momentum transfer $\tau$.

For the strict RFG model one assumes that an on-shell nucleon moving
in the initial-state momentum distribution is struck by the virtual
photon and an on-shell baryon resonance of mass $m_*$ is produced. Of
course this is a very over-simplified model (the struck nucleons are
not on-shell, the single-nucleon process is not all, the final state
is not a stable baryon but a decaying resonant or non-resonant state,
etc.), although it does provide some insight into a possible scaling
violation mechanism.  In particular it can shed some light on when
such scale breakings do or do not appear to be very large; that is, it
helps to define characteristic kinematic conditions where the QE-type
scaling could be expected to evolve into another type of scaling
behavior. 

Specifically, using the $N\to\Delta$ RFG model of \cite{ama99} and
forming the ratio $\rho_{12}$ defined above one typically obtains results above
unity in the $\psi^\prime$ region lying between the quasielastic and
``$\Delta$'' peaks. This is to be compared with the experimental ratio,
shown in Fig.~\ref{fig:r12slac}. For the upper panel in the figure the
$N\to\Delta$ RFG peak occurs at $\psi^\prime\cong 1.3$ for carbon (1.2
for gold), while for the lower panel the peak occurs at
$\psi^\prime\cong 0.72$ for carbon (0.67 for gold). The model yields a
typical ratio $\rho_{Au/C}$ in the region between the QE and
$N\to\Delta$ peaks of approximately 1.08.
For the conditions of the lower panel the model yields a ratio that is
closer to unity, approximately within 4--5\% of unity
in the region between the QE
and $N\to\Delta$ peaks. Thus we see, even with this over-simplified
model, that the remaining few percent deviation from unity of the
ratio $\rho_{12}$ is reasonably compatible with resonance production
in the $\psi^\prime>0$ region corresponding to excitation of the
$\Delta$.
 
Some further insight can be gained about why the ratio behaves as it
does by examining some of the specifics of the modeling.
In fact, in \cite{ama99} a scaling variable emerged naturally from
consideration of the RFG model for the ``$\Delta$'' region. Here we
generalize this to the scaling variable that occurs in this model for
inelastic single-nucleon transitions ($m_N\to m_*$).
With
\be
\rho_* \equiv 1 + \frac{\beta_*}{\tau} ,
\label{eq:rho-Delta}
\ee
where $\beta_*\equiv (m_*^2 - m_N^2)/4m_N^2\ge 0$, we have the analog of
Eq.~(\ref{eq:psi}):
\be
\psi_* \equiv \difrac{1}{\sqrt{\xi_F}}
\difrac
{\lambda-\tau\rho_*}
{ 
\sqrt{ (1+\lambda\rho_*)\tau + \kappa \sqrt{ \tau(\tau\rho_*^2+1) } } 
} .
\label{eq:psigen}
\ee
Futhermore, following the discussions in Sec.~\ref{sec:form}, 
one may shift the energy to obtain the analog of Eq.~(\ref{eq:psiprime}),
\be
\psi^\prime_* \equiv \difrac{1}{\sqrt{\xi_F}}
\difrac
{\lambda^\prime -\tau^\prime \rho_*^\prime }
{ 
\sqrt{ (1+\lambda^\prime \rho_*^\prime )\tau^\prime 
+ \kappa \sqrt{ \tau^\prime (\tau^\prime {\rho_*^\prime}^2+1) } } 
} ,
\label{eq:psiprimegen}
\ee
where $\rho_*' \equiv 1 + \beta_*/\tau'$.
If $m_*=m_N$, then $\beta_*=0$, which implies that $\rho_*=1$ and the
above generalized scaling variable yields the standard results in
Eqs.~(\ref{eq:psi}) and (\ref{eq:psiprime}). 
The RFG model for inelastic excitation of a stable
resonance of mass $m_*$ has scaling behavior of both the first and
second kinds in the variable $\psi_*$. In particular, if a {\em universal}
transverse scaling function $f_T$ underlies both quasielastic
scattering and the
excitation of some given resonance from nucleons in the nucleus, 
then we expect the former to occur in the total transverse cross section as
$f_T(\psi^\prime)$ weighted with the usual eN elastic cross section
and the latter as $f_T(\psi_*^\prime)$ weighted by the N$\to$N$^*$
inelastic cross section. Thus we expect (at least) that the scaling
behavior involves the way $f_T(\psi^\prime)$ varies with $\psi^\prime$
--- it scales --- versus the way $f_T(\psi_*^\prime)$ varies 
with $\psi^\prime$ --- note: not how it varies with $\psi_*^\prime$ in
which it scales.

Relationships can be established between $\psi_*$ and $\psi$ (or
${\psi_*}^\prime$ and ${\psi}^\prime$ using the primed 
kinematic variables as above --- in the following for simplicity we
consider only the unshifted variables):
\be
\psi_* = \sqrt{1+\gamma_*^2}\psi
  -\gamma_* \sqrt{\frac{2}{\xi_F}+\psi^2} ,
\label{eq:psirel}
\ee
where $2\gamma_*^2 \equiv [(1+\tau)(1+\rho_*^2\tau)]^{1/2}-(1+\rho_*\tau)$.
Substituting for $\rho_*$ one can write
\bea
\gamma_* &=& \frac{\beta_*}{\sqrt{2\tau (1+\tau)}}
  \left[ \sqrt{\left(1+\frac{\beta_*}{1+\tau} \right)^2
  +\frac{1}{\tau}\left(\frac{\beta_*}{1+\tau}\right)^2} +1+\frac{\beta_*}{1+\tau}
   \right]^{-1/2} \\
&\to& \frac{\beta_*}{2\sqrt{\tau (1+\tau)}} ,
\label{eq:gamma}
\eea
where the limiting behavior is for $\tau \gg 1$ for fixed $\beta_*$ 
({\it i.e.}, fixed
$m_*$). Under these conditions $\gamma_*$ falls as $[\tau
(1+\tau)]^{-1/2}$. Thus, if one keeps $m_*$ fixed, focusing on a
particular resonance, for instance, and examines the behavior of
$\psi_*$ as a function of $\tau$ and $\psi$ it is clear that as $\tau$
becomes very large $\gamma_*\to 0$ and therefore $\psi_*\to \psi$, so
that there will be no distinction between the two scaling variables.
Indeed, to get an idea of how large $\tau$ must be for this to occur,
suppose that in addition one has $|\eta_F \psi|\ll 1$, then
a simple approximation for $\psi_*$ is the following:
\be
\psi_* \cong \psi - \frac{\beta_*}{\eta_F \sqrt{\tau (1+\tau)}} .
\label{eq:gamma-appx}
\ee
So the value of $\tau$ that characterizes the coalescence of
$\psi_*$ and $\psi$ is
$\tau_0 = ([1+(2\beta_*/\eta_F)^2]^{1/2}-1)/2$. For instance, 
if $m_*=m_\Delta$ and $k_F=230$ MeV/c, then one has 
$\tau_0\cong 0.4$, corresponding to
$|Q_0^2|\cong 1.4\ ($GeV/c$)^2$. As the mass of the resonance grows
then so does the characteristic $\tau_0$, and the coalescence is
postponed until larger values of $\tau$.

Given these relationships between the various scaling variables, let
us now return to the analysis of 2$^{nd}$-kind scale breaking in the
resonance region. As noted above we expect some linear combination of 
$f_T(\psi^\prime)$ and $f_T(\psi_*^\prime)$ with weightings according
to the elastic and inelastic eN cross sections to be involved in the
region between the quasielastic peak ($\psi^\prime=0$) and the peak of
the specific resonance of interest ($\psi_*^\prime=0$). As a function
of $\psi^\prime$ the quasielastic contribution is assumed to 
exhibit 2$^{nd}$-kind scaling, whereas the N$\to$N$^*$ contribution
does not, because of the dependence on $k_F$ in the expressions
above. Consider for simplicity the result in
Eq.~(\ref{eq:gamma-appx}). With $\tau$ fixed and considering a given
value of $m_*$ and hence of $\beta_*$ we see that $\psi_*$ is less
than $\psi$ by the offset $\beta_*/[\eta_F\sqrt{\tau*(1+\tau)}]$; so
$\psi$ is some positive number whereas (being between the QE peak and
the resonance peak) $\psi_*$ is some negative number. For a small
$k_F$ the offset is large, whereas for a large $k_F$ it is small. That
is, for $k_F$ small $\psi_*$ is further away from zero than when $k_F$ is
large. Or, said a different way, the difference
\be
\psi_*(k_{F1})-\psi_*(k_{F2}) = \frac{\beta_*}{\sqrt{\tau*(1+\tau)}}
  \left[ \frac{1}{\eta_{F2}} - \frac{1}{\eta_{F1}} \right]
\label{eq:psistdiff}
\ee
is positive if $k_{F1}>k_{F2}$, and so $\psi_*(k_{F1})$ is closer to
zero in this case than is $\psi_*(k_{F2})$. Since the inelastic
response peaks when $\psi_*=0$, this implies that in the region
between the QE and N$\to$N$^*$ peaks the case with the larger $k_F$
should be {\em larger} than for the lower-$k_F$ case. Indeed this is
just what we observe from the data and, in fact, if the offset given
above is computed for typical conditions, the degree to which scaling
of the 2$^{nd}$-kind is broken is found to be quite compatible with
those results. These ideas have natural extensions to a distribution 
of resonances between each of which the same type of phenomenon will
occur. The actual magnitude of the effect of course depends on the
cross section weighting of each contribution and thus the present
arguments should only be regarded as qualitative ones --- a more
quantitative approach is presently being pursued. 

In summary, in effect, the kinematic offset from the QE scaling
variable which occurs in the scaling variable 
that enters naturally when considering baryon
excitation from nucleons in the nucleus alone appears to be capable of
explaining most of the breaking of 2$^{nd}$-kind scaling in the
resonance region.

\subsection{In the DIS Region}
\label{subsec:Dis}
Turning now to the DIS region, let us recall the conventional 
analysis of $x$-scaling 
(see, for example, \cite{clo79}). Noting that the quantity $\nu$
employed in DIS analyses is called $\omega$ in our work, as is usual,
and employing the above dimensionless kinematic variables we can write
\be
x = \frac{\tau}{\lambda}
\label{eq:xvar}
\ee
and see that when $\psi<0$ we have $x>1$ and the reverse; when $\psi=0$ we have 
$x=1$. 

Using the fact that $x=\tau/\lambda$ one can construct expressions
that directly relate $x$ to the scaling variable $\psi$ and Fermi
momentum (via $\xi_F$) --- in particular one useful form is the
following:
\be
x=\frac{2}{\xi_F}\left[ \left( \psi+\sqrt{\frac{2}{\xi_F}+\psi^2}
  \right)^2
  +\left\{\sqrt{1+\frac{1}{\tau}}-1\right\}2\psi
  \sqrt{\frac{2}{\xi_F}+\psi^2} \right]^{-1} .
\label{eq:xbj}
\ee
Or alternatively, if one wishes to work with the Nachtman variable
$\xi$ (not to be confused with $\xi_F$ used in this and other
RFG-motivated studies), then again various useful relationships exist.
In particular,
\bea
\xi &=& 2 (\kappa-\lambda) \\
&=& \frac{\xi_F}{1+\sqrt{1+\frac{1}{\tau}}} \left[
  \sqrt{\frac{2}{\xi_F} + \psi^2}-\psi \right]^2 .
\label{eq:nacht}
\eea
Moreover, it is possible to incorporate the energy
shift $E_{shift}$ as above by replacing all quantities on the
right-hand sides of Eqs.~(\ref{eq:xbj}--\ref{eq:nacht}) by their
primed counterparts to obtain shifted variables $x^\prime$ and
$\xi^\prime$. 

Let us examine the $x$-variable a little more closely. As written in
Eq.~(\ref{eq:xbj}) it is a function of $\tau$, $\xi_F$ (and hence
$k_F$) and $\psi$. The last two enter only in the combination $z\equiv
\sqrt{\xi_F/2}\psi\cong \eta_F \psi/2$, since
\be
\frac{1}{x} = \left[ \sqrt{1+z^2}+z \right]^2 +
\left\{\sqrt{1+\frac{1}{\tau}}-1\right\}
2 z \sqrt{1+z^2} .
\label{eq:xinv}
\ee
Thus, as $\tau\to\infty$ one finds that
\be
\frac{1}{x} \to \frac{1}{x^\infty} \equiv \left[ \sqrt{1+z^2}+z
\right]^2 .
\label{eq:xinfty}
\ee
This behavior holds for $\tau\gg\tau_{crit}$, where,
by comparing $x$ with $x^\infty$ in the regime where $\psi>0$, one finds that
\be
\tau_{crit} = \frac{z\sqrt{1+z^2}}{\left[\sqrt{1+z^2}+z\right]^2} 
\leq 1/4. 
\label{eq:taucrit}
\ee
That is, for $|Q^2|\gg 1$ (GeV/c)$^2$ one has that $x$ is well
approximated by $x^\infty$. In this regime $x$ becomes a function only
of $z$ and thus DIS scaling in $x$ ({\it i.e.},
independence of $\tau$) and 1$^{st}$-kind scaling in $\psi$ will occur
together. Furthermore, the above expression for $x^\infty$ yields
\bea
\frac{1}{x^\infty} &\cong& 1+2z \qquad {\rm for\ }\psi\ll
  \sqrt{2}/\eta_F \label{eq:breaklow}\\
&\cong& 4z^2 \ \qquad\quad {\rm for\ }\psi\gg \sqrt{2}/\eta_F .
\label{eq:breakhigh}
\eea
The ``asymptotic regime'' then occurs when $|Q^2|\gg 1$ (GeV/c)$^2$
and $x$ is small enough that $\psi\gg \sqrt{2}/\eta_F$ (typically
$\cong 5$--6), and under these conditions
\be
x \rightarrow \frac{1}{\left( \eta_F \psi \right)^2} .
\label{eq:xappx}
\ee
Thus in the asymptotic regime $x\propto 1/k_F^2$.

Re-expressing what we have seen in the analyses so far: 
throughout the region $x>1$, near $x=1$ (essentially the QE peak) and
down to about $x=0.3$--0.4 (corresponding to $\psi^\prime = 3$--4 in 
Fig.~\ref{fig:r12slac}) we find that the ratio $\rho_{12}$ is close to
unity and that its deviation can be reasonably accounted for by the
modest second-kind scale-breaking expected for resonance
excitation. Now we wish to explore the higher-$|Q^2|$, small-$x$
region in which DIS is expected to be the dominant reaction mechanism.

We begin by providing some ``translations'' between the nomenclature
and conventions employed for most studies done within the context of
nuclear physics and those done in particle physics studies.
Specifically, what are usually called $\sigma_L$ and $\sigma_T$ 
in the latter are given in terms of the 
responses $W_{1,2}$ and hence via Eqs.~(\ref{eq:RLT-W12}) of $R_{L,T}$
by the following (here $K$ is an arbitrary factor
conventionally taken to be $q$ or sometimes $\omega-\omega(\psi=0)$):
\bea
\sigma_T &=& \difrac{4 \pi^2 \alpha}{K}W_1                  
\label{eq:sigmaT-DIS}
\\
\sigma_L &=& \difrac{4 \pi^2 \alpha}{K}
\left[ \difrac{\kappa^2}{\tau}W_2 - W_1\right]  .  
\label{eq:sigmaL-DIS}
\eea
Their ratio $R$ is then
\be
\\
R \equiv \difrac{\sigma_L}{\sigma_T}=
\difrac{W_2}{W_1}\left[1 +\difrac{\lambda^2}{\tau}\right] -1 ,
\label{eq:R-DIS}
\ee
not to be confused with a different ratio
\be
{\cal R}\equiv \difrac{R_L}{R_T} = \difrac{\kappa^2}{2\tau}R .
\label{eq:calR-DIS}
\ee

\begin{figure}[htb]
\centerline{\mbox{\epsfysize=18cm \epsffile{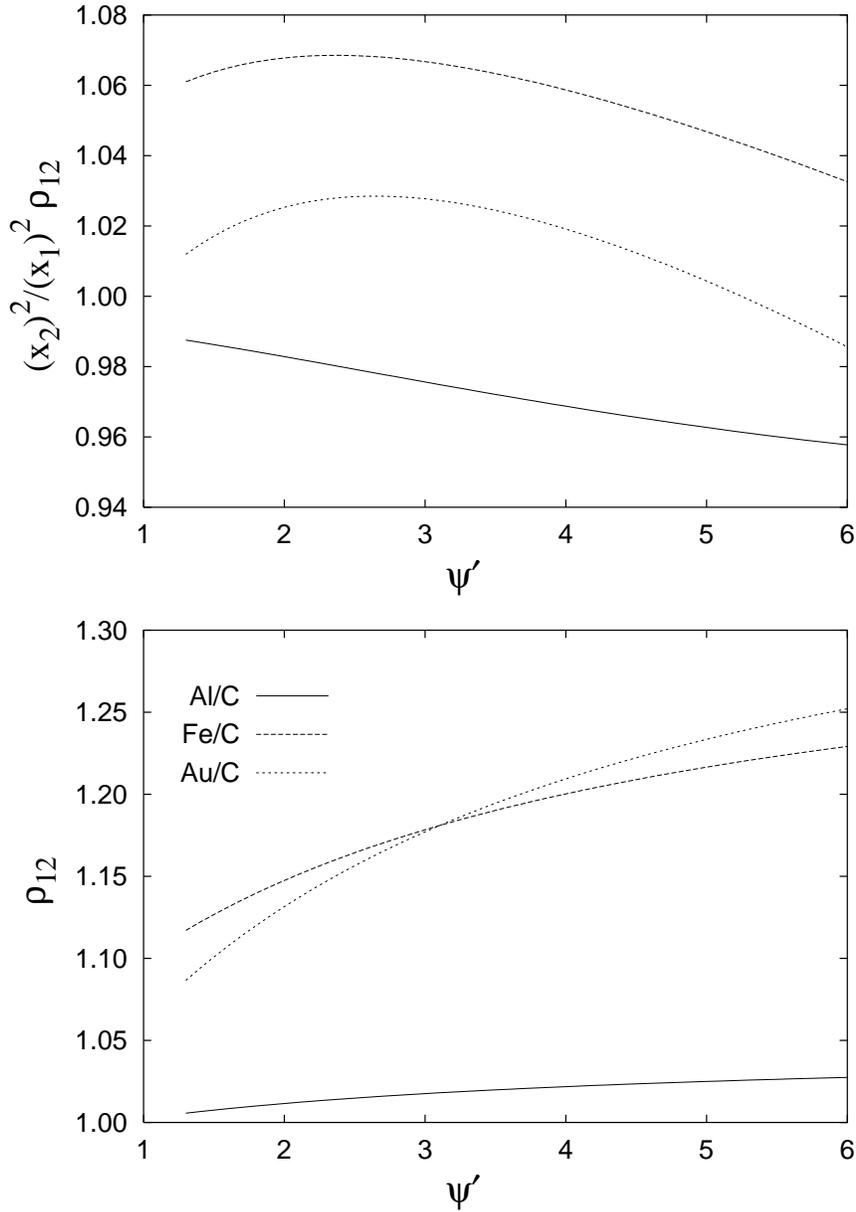}}}
\begin{center} 
\parbox{15cm}
{ \caption{ \label{fig:r12dis}  The ratio $\rho_{12}$ defined in the
    text for $|Q^2|=5$ (GeV/c)$^2$ and the same ratio multiplied by
    $(x_2)^2/(x_1)^2$ for pairs of nuclei 1 and 2.
}}   
\end{center}
\end{figure}

As in the previous discussions, we wish, to the degree that we can, 
to isolate the transverse part
of the inclusive cross section. In the DIS regime we must depend on
the limited information available for the ratio $R$ defined in 
Eq.~(\ref{eq:R-DIS}) \cite{das88,dasumore}. We then construct
\be
\Sigma_T=\frac{\sigma}{\left[ 1 + \frac{v_L}{v_T}{\cal R} \right]}
  =\frac{\sigma}{\left[ 1 + \frac{v_L}{v_T} \frac{\kappa^2}{2\tau}R
    \right]}
\label{eq:sigtdis}
\ee
and from this the transverse scaling function $f_T$ and the ratio
$\rho_{12}$, as above. In addition to the results shown in
Fig.~\ref{fig:r12slac} which are not really in the DIS region, we show
in Fig.~\ref{fig:r12dis} some of the (limited) information available at
higher-$|Q^2|$ --- apparently no information on individual eA cross
sections is available at very high momentum transfers and only the
usual EMC ratio at constant $x$ has been measured.

Scaling in the DIS regime is usually discussed in terms of the
quantities $F_1\equiv m_N W_1$ and $F_2\equiv \nu W_2$. 
The so-called EMC ratio is chosen to be $[F_2/A]_2/[F_2/A]_1$, 
where, as above, 1 and 2
label a given pair of nuclei. When this is plotted versus $x$ at very
high $|Q^2|$ it is seen to be independent of $|Q^2|$ and to exhibit
the so-called EMC effect. Now our previous analyses and the above
discussions show that if one chooses for the abscissa on a plot
to use the scaling variable $\psi^\prime$, {\em scaled by the 
Fermi momentum}, then one should also incorporate the charge of
variables into the ordinate in the plot by multiplying by
$k_F$. Explicitly this was the procedure used with $\psi^\prime\cong
y/k_F$ (see Eqs.~(\ref{eq:psiappx},\ref{eq:psiprime})) 
and $f\equiv k_F \times F$ (see Eqs.~(\ref{eq:ftotal},\ref{eq:fLT})).
Since here we always display results as functions of $\psi^\prime$, we
are then motivated also to employ a modified ``EMC-type'' ratio involving 
the structure functions for the two nuclear species 1 and 2, but now
with factors of $k_F$ for each:
\be
\rho^{DIS}_{12} \equiv 
\frac{\left[ k_{F}F_{2}/A\right] _{2}}{\left[ k_{F}F_{2}/A\right] _{1}} .
\label{ eq:rho-F2-DIS}
\ee

Let us now provide the connections between the QE-type ratio and the 
EMC-type ratio. These may be related via
\be
\rho^{DIS}_{12} \equiv  X_{12} \rho_{12} ,
\label{eq:rho-F2-DIS-a}
\ee
where the factor $X_{12}$ in turn may be decomposed into three
factors,
\be
X_{12} = R_{12} G_{12} K_{12} .
\label{eq:XQEDIS}
\ee
The first of these, 
\be
R_{12} = \difrac{[1+R]_2}{[1+R]_1} ,
\label{eq:ratRDIS}
\ee
arises simply because different parts of
the total cross section are conventionally chosen in forming the two
ratios. When comparing two different nuclear species {\em at constant}
$\psi^\prime$ it must be remembered that $x$ for species 1 is not the
same as $x$ for species 2, but that the two $x$-values are slightly
different. In fact, $R_{12}$ is quite close to unity for the range of
kinematics under study in this work. The second factor,
\be
G_{12}  = \difrac{[G_D^2]_2}{[G_D^2]_1} ,
\label{eq:ratGD}
\ee
stems from the fact that the QE ratio is defined in Eq.~(\ref{eq:rho})
via Eq.(\ref{eq:xi}) to have the
typical dipole form factor behavior expected for the transverse part
of the elastic eN cross section, whereas the DIS ratio is not --- in
the latter we expect e-quark scattering to begin to dominate. Again,
the fact that we display results versus $\psi^\prime$ for experiments
where incident electron energy $E_e$ and scattering angle $\theta_e$
are fixed means that for the two nuclear species the 4-momentum
transfers are slightly different. Indeed, $G_{12}$ is typically quite
close to unity for the range of kinematics under study. Note that, if
sufficient information were available to allow one to specify $|Q^2|$,
rather than have it vary as here (excepting the case shown in 
Fig.~\ref{fig:r12dis}), then this factor would be
irrelevant.

The important factor for our purposes is the third one,  
\be
K_{12} = \difrac{\left[ \lambda\tau^2/\kappa^3 \right]_2}
{\left[ \lambda\tau^2/\kappa^3 \right]_1} ,
\label{eq:ratxfact}
\ee
which contains all of the kinematic behavior involved in relating the
two types of ratios. At the quasielastic peak this is unity
(for simplicity $E_{shift}$ is neglected in these arguments), whereas
above the peak it in general differs from unity and so must
typically be taken into account. To get some feeling for how it enters
in the extreme DIS regime where $\tau\to\infty$ and $x\ll 1$,
note that there one has
\be
K_{12}^{\infty} = \difrac{\left[x^2\right]_2}
{\left[x^2\right]_1} .
\label{eq:xratio}
\ee
Thus in extreme-DIS kinematics the ratio defined in the present 
work in Eq.~(\ref{eq:rho}) for 
use in QE scattering differs from the conventional 
EMC-type ratio essentially by the factor
$\left[x^2\right]_2/\left[x^2\right]_1$ in Eq.~(\ref{eq:xratio}). From
Eq.~(\ref{eq:xappx}) above we see that this immediately implies that in the
asymptotic regime
\be
K_{12}^{\infty} \cong \difrac{\left[k_F^4\right]_1}
{\left[k_F^4\right]_2} .
\label{eq:xratiokf}
\ee

Returning to Fig.~\ref{fig:r12dis}, we see from the lower
panel that scaling of the second kind in this kinematic region is not
as good as it was for the resonance region and below.  
Multiplying $\rho_{12}$ by the ratio of the $x$'s as in the
upper panel in the figure yields results much closer to unity, as
expected. Namely, the upper panel corresponds to the EMC behavior when
results are displayed versus $\psi^\prime$. Indeed, uncertainties are
not shown in the figure as they are hard to quantify; what is clear is
that the results in the upper panel are essentially consistent with
only small deviations from unity. Finally, note that from the above
arguments asymptotically we would expect factors of $\left[k_F^4\right]_1/
\left[k_F^4\right]_2$, which from Table~1 are 1.15, 1.25 and 1.33
for Al/C, Fe/C and Au/C, respectively. While not unreasonable in fact
these are somewhat further from unity than the actual ratios in the
lower panel in the figure. Indeed, while $|Q^2|$ is reasonably 
large here ($\tau\cong 1.42>\tau_{crit}$ in Eq.~(\ref{eq:taucrit})), $x$
is not very small, ranging as it does from roughly 0.7 at the
left-hand side of the figure down to approximately only 0.25 at the
right-hand side. Or, in other words, the asymptotic regime where
$\psi^\prime$ is sufficiently large for Eq.~(\ref{eq:breakhigh}) to be valid
is only barely reached at the right-hand side of Fig.~\ref{fig:r12dis}
and the behavior seen here reflects the fact that these data lie
somewhere between the quasielastic and asymptotic regimes.

\section{Discussion and Conclusions}
\label{sec:concl}

In undertaking this study we have begun by refining our previous
analysis of the kinematic region lying below the quasielastic peak. We
had seen in our past work that scaling of the second kind is excellent
in this region and that values of $k_F$ and $E_{shift}$ can be
determined for each nuclear species for which medium- to high-energy
data exist. In the present work we have taken these determinations a
step further and evaluated how sensitive the scaling function is to
variations around the best-fit values of $k_F$ and $E_{shift}$,
finding that the former is very well determined (relative to an
overall multiplicative factor), whereas the latter is less important,
as found in our previous studies.

We have then employed the hypothesis of longitudinal superscaling to
extract the longitudinal contributions from the total inclusive
response and so obtain the transverse response and scaling
function. It should be stressed that the
kinematic regime in which the superscaling of $f_L$ is tested
experimentally is unfortunately rather limited and hence the
extraction of $f_T$ over a wide range of
kinematics is contingent on having $f_L$ superscale. However, it is
important to note that especially at high $q$, where the L/T ratio
becomes small, the cross section is known to be dominated by the transverse
response. Our focus in the present work is this region and thus the 
extraction procedure should be reasonably good here. 
Proceeding with this as a basic assumption using the $f_L$ determined
by what TL-separated data can be relied upon, we deduce $f_T$ for an
extended range of kinematics. We expect that breaking of scaling of
both the first and second kinds will be more prominent in $f_T$ than
in $f_L$ due to the well-known reaction mechanisms --- for instance,
pion production, $N\to\Delta$ excitations and 2-body meson-exchange
current effects, all of which are expected to be transverse-dominated
--- and accordingly have placed our focus on $f_T$. This is not to say
that $f_L$ cannot also have interesting scaling-breaking behavior ---
for instance, final-state interaction effects such as those arising
in the random-phase approximation can differ in
longitudinal and transverse, isoscalar and isovector channels ---
however, isolating $f_L$ except as we do here via the universal
superscaling approach has proven to be very difficult and thus the
focus on $f_T$ is inevitable at present.

We have selected two regimes for the present extended study, (1) the
resonance region above the quasielastic peak where we expect
$N\to\Delta$ and $N\to N^*$ contributions to be important, and (2)
still higher inelasticity where DIS takes over. Unfortunately, there
are very few cases for the latter in which eA cross sections are available for a
range of nuclei; most data involve ratios involving a pair of nuclei
for $x$ held constant, not what we want, namely, the individual cross
sections as functions of $|Q^2|$ and scaling variable $\psi^\prime$.

What we observe is that scaling of the first kind is clearly violated
when one proceeds above the QE peak ($\psi^\prime>0 \leftrightarrow
x<1$), whereas scaling of the second kind is better. Indeed, for
sufficiently high momentum transfers it is
excellent for $\psi^\prime<0 \leftrightarrow x>1$ and typically is
broken only by 10--15\% up through the resonance region. Moreover,
from simple modeling and even from simple kinematic arguments
involving the introduction of the scaling variable $\psi^\prime_*$
that occurs naturally when discussing the electroexcitation of
resonances, this modest second-kind scale-breaking is reasonably
accounted for, with very little further scaling violation left to
explain. This does not leave very much room for the other reaction mechanisms
that are known to violate scaling of the second kind. For instance, we
know that at least some of the meson-exchange effects depend on $k_F$
very differently than does the usual 1-body QE contribution (typically
in a way that breaks scaling of the second kind by contributions
proportional to
$k_F^3$), and thus the absence of significant contributions with this
behavior limits how much of a role they can play. In only a very
limited number of cases has it been possible to carry through modeling
of MEC effects with the requisite relativistic content and at present
a concerted effort is being made to extend past non-relativistic
treatments of MEC and interaction effects to the kinematic regime of
interest and thereby to test these ideas in more depth.

Finally, in the DIS region we see that there is clearly more breaking of
scaling of the second kind, indicating (as expected) that the reaction
mechanism is becoming different, presumably e-quark physics rather
than eN physics. The cross-over from the resonance region, where
second-kind scale-breaking is reasonably small and can be explained
using simple arguments, to the DIS region, where such does not appear to
be the case, may in fact be a relatively clear indicator of the shift
from hadronic to QCD degrees of freedom.

\vspace{0.5in}
{\Large {\bf Acknowledgements} }

\vspace{0.3in}
This work was supported in part by funds provided by the U.S. Department
of Energy under cooperative research agreement
\#DE-FC02-94ER40818, and by the Swiss National Science 
Foundation.  Additionally, 
CM wishes to express her thanks for the Bruno 
Rossi INFN-CTP Fellowship supporting 
her work while at M.I.T. and TWD/IS wish to thank the Institute 
for Nuclear Theory at 
the University of Washington for the hospitality while some 
of this work was being undertaken.

\vspace{0.3in}
MIT/CTP\#3168


\newpage
\begin{thebibliography}{999}
%
\bibitem{don99} T.W. Donnelly and I. Sick, 
Phys. Rev. Lett. {\bf 82}, 3212 (1999).
%
\bibitem{don99a} T.W. Donnelly and I. Sick, 
Phys. Rev. {\bf C60}, 065502 (1999).
%
\bibitem{Day90} D.B. Day, J.S. McCarthy, T.W. Donnelly and I. Sick,
Annu. Rev. Nucl. Part. Sci. {\bf 40}, 357 (1990).
%
\bibitem{alb88} W.M. Alberico, A. Molinari, T.W. Donnelly, 
L. Kronenberg and J.W. Van Orden,
Phys. Rev. {\bf C38}, 1801 (1988).
%
\bibitem{bar98} M.B. Barbaro, R. Cenni, A. DePace, T.W. Donnelly 
and A. Molinari, Nucl. Phys. {\bf A643}, 137 (1998).
%
\bibitem{cen97} R. Cenni, T.W. Donnelly and A. Molinari,
Phys. Rev. {\bf C56}, 276 (1997).
%
\bibitem{Whitney74}
R.R. Whitney, I.~Sick, J.R. Ficenec, R.D. Kephart, and W.P. Trower.
\newblock {\em Phys. Rev.} {\bf C9}, 2230 (1974).
%
\bibitem{Arrington99}
J.~Arrington, C.S. Armstrong, T.~Averett, O.~Baker, L.~deBever, C.~Bochna,
  W.~Boeglin, B.~Bray, R.~Carlini, G.~Collins, C.~Cothran, D.~Crabb, D.~Day,
  J.~Dunne, D.~Dutta, R.~Ent, B.~Fillipone, A.~Honegger, E.~Hughes, J.~Jensen,
  J.~Jourdan, C.~Keppel, D.~Koltenuk, R.~Lindgren, A.~Lung, D.~Mack,
  J.~McCarthy, R.~McKeown, D.~Meekins, J.~Mitchell, H.~Mkrtchyan, G.~Niculescu,
  T.~Petitjean, O.~Rondon, I.~Sick, C.~Smith, B.~Tesburg, W.~Vulcan, S.~Wood,
  C.~Yan, J.~Zhao, and B.~Zihlmann.
\newblock {\em Phys. Rev. Lett.} {\bf 82}, 2056 (1999).
%
\bibitem{Meziani92}
Z.-E. Meziani, J.P. Chen, D.~Beck, G.~Boyd, L.M. Chinitz, D.B. Day, L.C.
  Dennis, G.E. Dodge, B.W. Fillipone, K.L. Giovanetti, J.~Jourdan, K.W. Kemper,
  T.~Koh, W.~Lorenzon, J.S. McCarthy, R.D. McKeown, R.G.Milner, R.C. Minehart,
  J.~Morgenstern, J.~Mougey, D.H Potterveld, O.A. Rondon-Aramayo, R.M. Sealock,
  I.~Sick, L.C. Smith, S.T. Thornton, R.C. Walker, and C.~Woodward.
\newblock {\em Phys. Rev. Lett.} {\bf 69}, 41 (1992).
%
\bibitem{Sealock89}
R.M. Sealock, K.L. Giovanetti, S.T. Thornton, Z.-E. Meziani, O.A. Rondon-Aramayo,
  S.~Auffret, J.-P. Chen, D.G. Christian, D.B. Day, J.S. McCarthy, R.C.
  Minehard, L.C. Dennis, K.W. Kemper, B.A. Mecking, and J.~Morgenstern.
\newblock {\em Phys. Rev. Lett.} {\bf 62}, 1350 (1989).
%
\bibitem{Day93}
D.~Day, J.S. McCarthy, Z.-E. Meziani, R.~Minehart, R.~Sealock, S.T. Thornton,
  J.~Jourdan, I.~Sick, B.W. Filippone, R.D. McKeown, R.G. Milner, D.H.
  Potterveld, and Z.~Szalata.
\newblock {\em Phys. Rev.} {\bf C48}, 1849 (1993).
%
\bibitem{Rock82a}
S.~Rock, R.G. Arnold, B.T. Chertok, Z.M. Szalata, D.~Day, J.S. McCarthy,
  F.~Martin, B.A. Mecking, I.~Sick, and G.~Tamas.
\newblock {\em Phys. Rev.} {\bf C26}, 1592 (1982).
%
\bibitem{Barreau81}
P.~Barreau, M.~Bernheim, M.~Brussel, G.P. Capitani, J.~Duclos, J.M. Finn,
  S.~Frullani, F.~Garibaldi, D.~Isabelle, E.~Jans, J.~Morgenstern, J.~Mougey,
  D.~Royer, B.~Saghai, E.~de~Sanctis, I.~Sick, D.~Tarnowski, S.~Turck-Chieze,
  and P.D. Zimmermann.
\newblock {\em Nucl. Phys.} {\bf A358}, 287 (1981).
%
\bibitem{Barreau83}
P.~Barreau, M.~Bernheim, J.~Duclos, J.M. Finn, Z.~Meziani, J.~Morgenstern,
  J.~Mougey, D.~Royer, B.~Saghai, D.~Tarnowski, S.~Turck-Chieze, M.~Brussel,
  G.P. Capitani, E.~de~Sanctis, S.~Frullani, F.~Garibaldi, D.B. Isabelle,
  E.~Jans, I.~Sick, and P.D. Zimmermann.
\newblock {\em Nucl. Phys.} {\bf A402}, 515 (1983).
%
\bibitem{Baran88}
D.T. Baran, B.F. Filippone, D.~Geesaman, M. Green, R.J. Holt, H.E. Jackson,
  J.~Jourdan, R.D. McKeown, R.G. Milner, J.~Morgenstern, D.H. Potterveld, R.E.
  Segel, P.~Seidl, R.C. Walker, and B.~Zeidman.
\newblock {\em Phys. Rev. Lett.} {\bf 61}, 400 (1988).
%
\bibitem{Connell87}
J.S. O'Connell, W.R. Dodge, Jr. J.W.~Lightbody, X.K. Maruyama, J.O. Adler,
  K.~Hansen, B.~Schroeder, A.M. Bernstein, K.I. Blomqvist, B.H. Cottman, J.J.
  Comuzzi, R.A. Miskimen, B.P. Quinn, J.H. Koch, and N.~Ohtsuka.
\newblock {\em Phys. Rev.} {\bf C35}, 1063 (1987).
%
\bibitem{Deady86}
M.~Deady, C.F. Williamson, J.~Wong, P.D. Zimmerman, C.~Blatchley, J.M. Finn,
  J.~LeRose, P.~Sioshans, R.~Altemus, J.S. McCarthy, and R.R. Whitney.
\newblock {\em Phys. Rev.} {\bf C33}, 1897 (1986).
%
\bibitem{Meziani85}
Z.~Meziani, P.~Barreau, M.~Bernheim, J.~Morgenstern, S.~Turck-Chieze,
  R.~Altemus, J.~McCarthy, L.J. Orphanos, R.R. Whitney, G.P. Capitani,
  E.~DeSanctis, S.~Frullani, and F.~Garibaldi.
\newblock {\em Phys. Rev. Lett.} {\bf 54}, 1233 (1985).
%
\bibitem{Yates93}
T.C. Yates, C.F. Williamson, W.M. Schmitt, M.~Osborn, M.~Deady, P.~Zimmerman,
  C.C. Blatchley, K.~Seth, M.~Sarmiento, B.~Barker, Y.~Jin, L.E. Wright, and
  D.S. Onley.
\newblock {\em Phys. Lett.} {\bf B312}, 382 (1993).
%
\bibitem{Williamson97}
C.F. Williamson, T.C. Yates, W.M. Schmitt, M.~Osborn, M.~Deady, P.D. Zimmerman,
  C.C. Blatchley, K.K. Seth, M.~Sarmiento, B.~Parker, Y.~Jin, L.E. Wright, and
  D.S. Onley.
\newblock {\em Phys. Rev.} {\bf C56}, 3152 (1997).
%
\bibitem{Hotta84}
A.~Hotta, P.J. Ryan, H.~Ogino, B.~Parker, G.A. Peterson, and R.P. Singhal.
\newblock {\em Phys. Rev.} {\bf C30}, 87 (1984).
%
\bibitem{Chen91}
J.P. Chen, Z.-E. Meziani, G.~Boyd, L.M. Chinitz, D.B. Day, L.C. Dennis,
  G.~Dodge, B.W. Filippone, K.L. Giovanetti, J.~Jourdan, K.W. Kemper, T.~Koh,
  W.~Lorenzon, J.S. McCarthy, R.D. McKeown, R.G. Milner, R.C. Minehart,
  J.~Morgenstern, J.~Mougey, D.H. Potterveld, O.A. Rondon-Aramayo, R.M.Sealock,
  L.C. Smith, S.T. Thornton, R.C. Walker, and C.~Woodward.
\newblock {\em Phys. Rev. Lett.} {\bf 66}, 1283 (1991).
%
\bibitem{Blatchley86}
C.C. Blatchley, J.J. LeRose, O.E. Pruet, P.D. Zimmerman, C.F. Williamson, and
  M.~Deady.
\newblock {\em Phys. Rev.} {\bf C34}, 1243 (1986).
%
\bibitem{jour96} J. Jourdan, 
Nucl. Phys. {\bf A603}, 117 (1996).
%
\bibitem{ama01} J.E. Amaro, M.B. Barbaro, J.A. Caballero, 
T.W. Donnelly and A. Molinari, 
to be published in Nucl. Phys. A.
%
\bibitem{ama01a} J.E. Amaro, M.B. Barbaro, J.A. Caballero, 
T.W. Donnelly and A. Molinari, 
submitted to Nucl. Phys. A.
%
\bibitem{car01} J. Carlson, J. Jourdan, R. Schiavilla and
I.~Sick, submitted to Phys. Rev. C.
%
\bibitem{van81} J.W. Van Orden and T.W. Donnelly,
Ann. Phys. {\bf 131}, 451 (1981).
%
\bibitem{ama99} J.E. Amaro, M.B. Barbaro, J.A. Caballero, 
T.W. Donnelly and A. Molinari, 
Nucl. Phys. {\bf A657}, 161 (1999).
%
\bibitem{clo79} F.E. Close, {\it An Introduction to Quarks and Partons,}
Academic Press (1979).
%
\bibitem{das88} S.R. Dasu, ``Precision
Measurement of $x$, $Q^2$ and $A$-dependence of $R=\sigma_L/\sigma_T$
and $F_2$ in Deep Inelastic Scattering'', UR-1059, ER13065-535,
Apr. 1988 (Ph.D. thesis, University of Rochester). 
%
\bibitem{dasumore} S. Dasu {\it et al.,} 
Phys. Rev. Lett. {\bf 60}, 2591 (1988); 
Phys. Rev. Lett. {\bf 61}, 1061 (1988); and
Phys. Rev. {\bf D49}, 5641 (1994).
%
\end{thebibliography}
%



%
\newpage

\newpage

\newpage

\newpage

\newpage

\newpage

\newpage

\newpage

\newpage


\end{document}